\documentclass[journal]{IEEEtran}

\usepackage{graphicx}
\usepackage{epstopdf}
\usepackage{standalone}
\usepackage[nolist ]{acronym}
\usepackage{tcolorbox}

\usepackage{pdfpages}
\usepackage{tikz}
\usetikzlibrary{positioning}
\usepackage{hyperref}
\usepackage{pdfcomment}
\usepackage{hologo}

\usepackage{xstring}

\usepackage[utf8]{inputenc}
\usepackage{marvosym}

\usepackage{algorithm}
\usepackage{algpseudocode}
\usepackage{tikz}

\usepackage{listings}
\definecolor{ListingBackground}{rgb}{0.97,0.97,0.97}
\usepackage{pgfplots}
\pgfplotsset{compat=newest}
\pgfplotsset{
	box plot/.style={
		/pgfplots/.cd,
		fill=blue!30,
		only marks,
		mark=-,
		mark size=0.2em,
		/pgfplots/error bars/.cd,
		y dir=plus,
		y explicit,
	},
	box plot box/.style={
		/pgfplots/error bars/draw error bar/.code 2 args={%
			\draw  ##1 -- ++(.2em,0pt) |- ##2 -- ++(-.2em,0pt) |- ##1 -- cycle;
		},
		/pgfplots/table/.cd,
		y index=2,
		y error expr={\thisrowno{3}-\thisrowno{2}},
		/pgfplots/box plot
	},
	box plot top whisker/.style={
		/pgfplots/error bars/draw error bar/.code 2 args={%
			\pgfkeysgetvalue{/pgfplots/error bars/error mark}%
			{\pgfplotserrorbarsmark}%
			\pgfkeysgetvalue{/pgfplots/error bars/error mark options}%
			{\pgfplotserrorbarsmarkopts}%
			\path ##1 -- ##2;
		},
		/pgfplots/table/.cd,
		y index=4,
		y error expr={\thisrowno{2}-\thisrowno{4}},
		/pgfplots/box plot
	},
	box plot bottom whisker/.style={
		/pgfplots/error bars/draw error bar/.code 2 args={%
			\pgfkeysgetvalue{/pgfplots/error bars/error mark}%
			{\pgfplotserrorbarsmark}%
			\pgfkeysgetvalue{/pgfplots/error bars/error mark options}%
			{\pgfplotserrorbarsmarkopts}%
			\path ##1 -- ##2;
		},
		/pgfplots/table/.cd,
		y index=5,
		y error expr={\thisrowno{3}-\thisrowno{5}},
		/pgfplots/box plot
	},
	box plot median/.style={
		/pgfplots/box plot
	},
	boxplot/every median/.style={
		ultra thick,dashed,cyan
	}
}

\definecolor{flexicolor}{RGB}{46,49,146}
\definecolor{amaricolor}{RGB}{237,28,36}

\usepackage{xspace}

\usepackage[binary-units=true]{siunitx}
\usepackage{psfrag}
\usepackage{graphicx}
\usepackage{tabularx,booktabs}
\usepackage{multirow}
\usepackage{rotating}

\renewcommand{\baselinestretch}{1}

\usepackage{multirow}

\usepackage[cmex10]{amsmath}
\usepackage[caption=false,font=footnotesize]{subfig}
\usepackage{stfloats}
\begin{document}
	
	\newcommand{\paperTitle}{Data-driven Network Simulation for Performance Analysis of Anticipatory Vehicular \\ Communication Systems}
\newcommand{\paperAuthors}{Benjamin Sliwa and Christian Wietfeld}
\newcommand{\paperEmails}{$\{$Benjamin.Sliwa, Christian.Wietfeld$\}$@tu-dortmund.de}

\newcommand{\figurePadding}{0pt}
\newcommand{\figureTopPadding}{\figurePadding}
\newcommand{\figureBottomPadding}{\figurePadding}

\newcommand\yRf{$\tilde{\mathbf{Y}}_{\text{RF}}$\xspace}
\newcommand\yGpr{$\tilde{\mathbf{Y}}_{\text{GPR}}$\xspace}

\newcommand\mYRf{\tilde{\mathbf{Y}}_{\text{RF}}}
\newcommand\mYGpr{\tilde{\mathbf{Y}}_{\text{GPR}}}

\renewcommand{\emph}{\textit}

\newcommand{\vs}
{
	\ifdefined\singleColumn
		-0.8cm
	\else
		-0.4cm
	\fi	
}

\newcommand{\takeAway}[2]
{\colorbox{backcolour}
	{
		\begin{minipage}{1\textwidth}
			\section*{\summary: #1}
			#2
		\end{minipage}
		}
}

\newcommand{\dummy}[3]
{
	\begin{figure}[b!]  
		\begin{tikzpicture}
		\node[draw,minimum height=6cm,minimum width=\columnwidth]{\LARGE #1};
		\end{tikzpicture}
		\caption{#2}
		\label{#3}
	\end{figure}
}

\newcommand{\wDummy}[3]
{
	\begin{figure*}[b!]  
		\begin{tikzpicture}
		\node[draw,minimum height=6cm,minimum width=\textwidth]{\LARGE #1};
		\end{tikzpicture}
		\caption{#2}
		\label{#3}
	\end{figure*}
}

\newcommand{\basicFig}[7]
{
	\begin{figure}[#1]  	
		\vspace{#6}
		\centering

		\ifdefined\singleColumn
			\includegraphics[width=0.6\columnwidth]{#2}
		\else
			\includegraphics[width=#7\columnwidth]{#2}
		\fi
		
		\caption{#3}
		\label{#4}
		\vspace{#5}	
	\end{figure}
}
\newcommand{\fig}[4]{\basicFig{#1}{#2}{#3}{#4}{0cm}{0cm}{1}}

\newcommand{\subfig}[3]
{
	\subfloat[#3]{\includegraphics[width=#2\textwidth]{#1}}\hfill
}

\newcommand\circled[1] 
{
	\tikz[baseline=(char.base)]
	{
		\node[shape=circle,draw,inner sep=1pt] (char) {#1};
	}\xspace
}

\newcommand{\sideHeader}[3]
{
	\multirow{#1}{*}{
		\rotatebox[origin=c]{90}{
			\parbox{#2}{\centering \textbf{#3}}
		}
	}
}

\newcommand\sldes{system-level \ac{DES}\xspace}
\newcommand\des{\ac{DES}\xspace}
\newcommand\ddns{\ac{DDNS}\xspace}
\newcommand\mno{\ac{MNO}\xspace}
\newcommand\mnos{\acp{MNO}\xspace}
\newcommand\ann{\ac{ANN}\xspace}
\newcommand\rf{\ac{RF}\xspace}
\newcommand\svm{\ac{SVM}\xspace}
\newcommand\enb{\ac{eNB}\xspace}
\newcommand\mnoA{\emph{\ac{MNO}~A}\xspace}
\newcommand\mnoB{\emph{\ac{MNO}~B}\xspace}
\newcommand\mnoC{\emph{\ac{MNO}~C}\xspace}

	\begin{acronym}
	\acro{HD}{High Definition}
	\acro{UAV}{Unmanned Aerial Vehicle}
	\acro{SUS}{System Under Study}
	\acro{AI}{Artificial Intelligence}
	\acro{CR}{Cognitive Radio}
	\acro{DDNS}{Data-driven Network Simulation}
	\acro{DES}{Discrete Event Simulation}
	\acro{MUS}{Method Under Study}
	\acro{TRUST}{Throughput prediction based on LSTM}
	\acro{LSTM}{Long Short-term Memory}
	\acro{CA}{Carrier Aggregation}
	\acro{ECDF}{Empirical Cumulative Distribution Function}
	\acro{QoS}{Quality of Service}
	\acro{URLLC}{Ultra Reliable Low Latency Communications}
	\acro{mMTC}{Massive Machine-type Communications}
	\acro{eMBB}{Enhanced Mobile Broadband}
	\acro{LIDAR}{Light Detection and Ranging}
	\acro{CAV}{Connected and Automated Vehicle}
	\acro{LR}{Linear Regression}
	\acro{TPC}{Transmission Power Control}
	
	\acro{TCP}{Transmission Control Protocol}
	\acro{KPI}{Key Performance Indicator}
	\acro{RTT}{Round Trip Time}		
	\acro{MTC}{Machine-type Communication}

	\acro{NIC}{Network Interface Card}
	\acro{PDCP}{Packet Data Convergence Protocol}
	\acro{RLC}{Radio Link Control}
	\acro{MAC}{Medium Access Control}
	\acro{HARQ}{Hybrid Automatic Repeat Request}
	\acro{IP}{Internet Protocol}

	\acro{ANN}{Artificial Neural Network}
	\acro{CART}{Classification And Regression Tree}
	\acro{GPR}{Gaussian Process Regression}
	\acro{M5}{M5 Regression Tree}
	\acro{RF}{Random Forest}
	\acro{SVM}{Support Vector Machine}
	\acro{SMO}{Sequential Minimal Optimization}
	\acro{RBF}{Radial Basis Function}
	\acro{WEKA}{Waikato Environment for Knowledge Analysis}
	\acro{MDI}{Mean Decrease Impurity}
	\acro{SGD}{Stochastic Gradient Descent}
	\acro{KNN}{k-Nearest Neighbors}
	\acro{LR}{Logistic Regression}
	
	\acro{CM}{Connectivity Map}
	\acro{CAT}{Channel-aware Transmission}
	\acro{ML-CAT}{Machine Learning CAT}
	\acro{pCAT}{predictive CAT}
	\acro{ML-pCAT}{Machine Learning pCAT}
		
	\acro{ITS}{Intelligent Transportation System}
	\acro{LIMoSim}{Lightweight ICT-centric Mobility Simulation}
	\acro{OMNeT++}{Objective Modular Network Testbed in C++}
	\acro{ns-3}{Network Simulator 3}
	
	\acro{RAIK}{Regional Analysis to Infer KPIs}
	\acro{RAT}{Radio Access Technology}
	\acro{MNO}{Mobile Network Operator}
	\acro{LTE}{Long Term Evolution}
	\acro{UE}{User Equipment}
	\acro{eNB}{evolved Node B}
	\acro{RSRP}{Reference Signal Received Power}
	\acro{RSRQ}{Reference Signal Received Quality}
	\acro{SINR}{Signal-to-interference-plus-noise Ratio}
	\acro{CQI}{Channel Quality Indicator}
	\acro{ASU}{Arbitrary Strength Unit} 
	\acro{TA}{Timing Advance}
	\acro{NWDAF}{Network Data Analytics Function}
\end{acronym}

	\acresetall
	\title{\paperTitle}

\author{Benjamin Sliwa \emph{Student Member, IEEE}, and Christian Wietfeld \emph{Senior Member, IEEE}
	\thanks{The authors are with Communication Networks Institute, TU Dortmund University, 44227 Dortmund, Germany
		{\tt\small $\{$Benjamin.Sliwa, Christian.Wietfeld$\}$@tu-dortmund.de}}%
}

\maketitle
	\begin{tikzpicture}[remember picture, overlay]
\node[below=5mm of current page.north, text width=20cm,font=\sffamily\footnotesize,align=center] {Published in: IEEE Access\\DOI: \href{http://dx.doi.org/10.1109/ACCESS.2019.2956211}{10.1109/ACCESS.2019.2956211}\vspace{0.3cm}\\\pdfcomment[color=yellow,icon=Note]{
@Article\{Sliwa/etal/2019d,\\
  author  = \{Benjamin Sliwa and Christian Wietfeld\},\\
  title   = \{Data-driven network simulation for performance analysis of anticipatory vehicular communication systems\},\\
  journal = \{IEEE Access\},\\
  year    = \{2019\},\\
  month   = \{Nov\},\\
\}
}};
\node[above=5mm of current page.south, text width=15cm,font=\sffamily\footnotesize] {2019~IEEE. Personal use of this material is permitted. Permission from IEEE must be obtained for all other uses, including reprinting/republishing this material for advertising or promotional purposes, collecting new collected works for resale or redistribution to servers or lists, or reuse of any copyrighted component of this work in other works.};
\end{tikzpicture}

	\begin{abstract}
%
%
The provision of reliable connectivity is envisioned as a key enabler for future autonomous driving. Anticipatory communication techniques have been proposed for proactively considering the properties of the highly dynamic radio channel within the communication systems themselves.
%
%
Since real world experiments are highly time-consuming and lack a controllable environment, performance evaluations and parameter studies for novel anticipatory vehicular communication systems are typically carried out based on network simulations. However, due to the required simplifications and the wide range of unknown parameters (e.g., \ac{MNO}-specific configurations of the network infrastructure), the achieved results often differ significantly from the behavior in real world evaluations.
%
%
In this paper, we present \ddns as a novel data-driven approach for analyzing and optimizing anticipatory vehicular communication systems. Different machine learning models are combined for achieving a close to reality representation of the analyzed system's behavior.
%
%
%
%
In a proof of concept evaluation focusing on opportunistic vehicular data transfer, the proposed method is validated against field measurements and system-level network simulation. In contrast to the latter, \ddns does not only provide massively faster result generation, it also achieves a significantly better representation of the real world behavior due to implicit consideration of cross-layer dependencies by the machine learning approach.
\end{abstract}


	\IEEEpeerreviewmaketitle
	
	\section{Introduction} \label{sec:introduction}

Within the approaching transition phase from human-driven cars to fully-autonomous traffic systems \cite{Sliwa/etal/2019b}, guaranteeing reliable and efficient communication is of crucial importance for enabling mutual coordination between the traffic participants as well as for optimizing the \ac{ITS}-based traffic flow by using the vehicles themselves as mobile sensors.
%
%
In order to provide seamless connectivity and avoid link failures proactively, future communication technologies will rely on short and mid term predictions of the radio channel quality and meaningful end-to-end indicators. Context-aware and \emph{anticipatory} \cite{Bui/etal/2017a} mobile networking principles such as opportunistic channel access \cite{Sliwa/etal/2018a} and dynamic \ac{RAT} selection \cite{Sepulcre/Gozalvez/2018a} have been demonstrated to be able to significantly improve the end-to-end \ac{QoS} of challenging data links.
%
%
In order to fulfill the requirements of upcoming 5G networks for \ac{URLLC}, \ac{mMTC}, and \ac{eMBB}, these methods need be brought to the next performance level. The exploitation of machine learning offers the potential to be the catalyst for this development \cite{Akpakwu/etal/2018a}, as its inherent strength is to leverage \emph{hidden interdependencies} between measurable variables, which are mostly too complex to be covered in an analytical solution.

%
%
The development process of these novel anticipatory vehicular communication systems confronts researchers and engineers  with a \emph{methodological dilemma}: While the most accurate estimations for the future real world performance can be achieved by performing real world experiments, this approach is highly time consuming and lacks a controllable environment. In fact, it is practically impossible to guarantee fairness by evaluating different methods under the exact same network conditions.
%
%
System-level network simulation based on \ac{DES} has emerged as the most commonly used scientific method to analyze mobile communication systems \cite{Cavalcanti/etal/2018a}, due to its capability of solving both issues. However, the necessary model simplifications reduce the significance of the achieved results for making conclusions about the real world behavior of the analyzed \ac{SUS}.
%
%
\fig{!b}{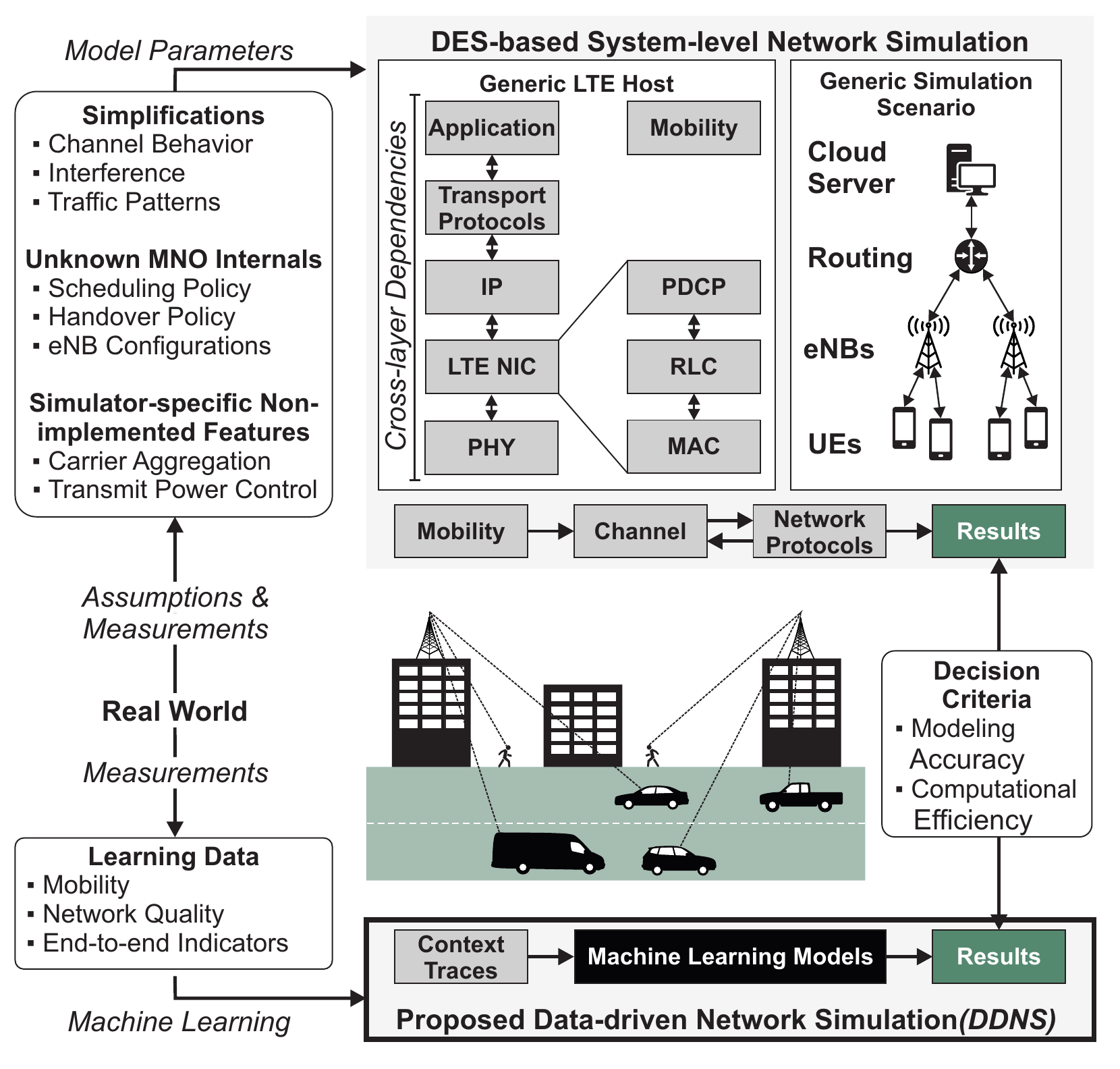}{Comparison of modeling complexity and implicated challenges for classical system-level network simulation and the proposed \ac{DDNS}.}{fig:DDNS}
%
%
%

%
%
In this paper, we present \ac{DDNS} as a novel approach for simulating the end-to-end behavior of vehicular communication networks. Through application of a data-driven approach and a combination of multiple machine learning models, the proposed method is able to achieve a level of accuracy almost similar to real world evaluations, the computational efficiency of analytical modeling and the environment control of classical network simulation. 

Fig.~\ref{fig:DDNS} shows a comparison of the modeling complexity between the proposed \ddns and classical \des (the architecture models are inspired by the implementations of the SimuLTE framework \cite{Virdis/etal/2015a}). As the \des approach involves a large amount of submodules on all logical layers, the model parameterization within the simulation setup phase is highly complex. Moreover, many of the required parameters are either subject to simplifications or are even unknown due to confidential \mno-specific configurations. 
In contrast to that, the proposed \ddns method focuses on direct modeling of the end-to-end behavior. The complex interdependencies between the different components are not explicitly parameterized. Instead, they are implicitly learned solely from the data within the training phase of the machine learning models.

%
%
This manuscript extends and brings together groundwork for data-driven network simulation \cite{Sliwa/Wietfeld/2019a}, data rate prediction \cite{Sliwa/Wietfeld/2019b} and anticipatory data transmission in vehicular networks \cite{Sliwa/etal/2018a, Sliwa/etal/2018b, Sliwa/etal/2019a}. In contrast to the previous work, we consider additional experiments, further machine learning methods and provide an extended theoretical discussion. Furthermore, all evaluations are performed in uplink and downlink transmission direction, whereas the previous work focused only on the uplink performance.
%
%
The contributions provided by this paper are summarized as follows:
\begin{itemize}
	\item Presentation of \textbf{\acf{DDNS}} as a novel performance analysis method for evaluating and optimizing the end-to-end behavior of anticipatory vehicular communication systems.
	\item Comparison of different machine learning approaches for \textbf{client-based online data rate prediction} in vehicular \ac{LTE} networks.
	\item \textbf{Validation against field measurements} and comparison to classical system-level network simulation in a proof of concept study focusing on opportunistic vehicular data transfer.
	\item All raw results and the developed applications are provided in an \textbf{open source} way.
\end{itemize}
%
%
The remainder of the paper is structured as follows. After discussing relevant related research in Sec.~\ref{sec:related}, we introduce methodological aspects in Sec.~\ref{sec:methods}. Afterwards, we present the machine learning-based solution approach for data rate prediction in vehicular multi-\mno networks in Sec.~\ref{sec:prediction}, which is a key component for the proposed \ddns method proposed in Sec.~\ref{sec:DDNS}. For the validation of the proposed approach, we consider a case study focusing on opportunistic vehicular data transfer in Sec.~\ref{sec:validation}. Finally, we summarize the key properties and the limitations of the \ac{DDNS} method in Sec.~\ref{sec:limitations}.

	\section{Related Work} \label{sec:related}

%
%

%
%
\textbf{Methods for network performance analysis}: 
Due to the complex interdependencies of mobility and communication, analysis and development of next generation \acp{CAV} and \acp{ITS} require the joint consideration of both domains \cite{Djahel/etal/2015a, Chen/etal/2017a}.
%
%
System-level network simulation has become the main evaluation method for vehicular communications systems, however the analysis carried out in \cite{Cavalcanti/etal/2018a} shows that a high number of publications rely on too simplistic parameter assumptions.
%
%
Although a lot of effort is spent on making these simulations more realistic \cite{Mir/2018a}, the underlying issues are often only shifted to a different domain. As a popular example, ray tracing-based analysis \cite{Yun/Iskander/2015a} theoretically allows to obtain detailed insights into the radio propagation characteristics within well-defined scenarios. However, the required environment data -- highly detailed maps with obstacle shape and material information -- is often not available. In addition, increasing the level of detail within those simulations inherently increases the computation time and therefore limits its applicability for large-scale evaluations.

%
%
\textbf{Machine learning}: 
The application of machine learning methods offers new potentials for modeling and analyzing mobile wireless communication systems. While analytical models fail to consider the complex interdependencies between the considered variables in highly dynamic environments, those impacts can be implicitly learned by machine learning-based models. Giordani et al. \cite{Giordani/etal/2019a} even envision future 6G networks to bring intelligence to every terminal in the network.
%
%
A general summary about machine learning methods and their application fields within wireless communication networks is provided by \cite{Jiang/etal/2017a}. In addition, Ye et al. \cite{Ye/etal/2018a} and Liang et al. \cite{Liang/etal/2019a} present summaries with a deeper focus on vehicular networks. 
%
%
Recently, the idea of learning the end-to-end behavior of communication systems has received great attention within the wireless communications community \cite{Qin/etal/2019a}. First approaches, which focus on learning the physical layer behavior, have been proposed by Ye et al. \cite{Ye/etal/2018b}, D\"orner et al. \cite{Doerner/etal/2018a}, and Aoudia et al. \cite{Aoudia/Hoydis/2018a}. By interpreting the communication system as an \emph{autoencoder}, the behavior can be learned in a supervised manner based on \ac{SGD} without requiring channel models for the physical layer interactions. The work presented in this manuscript can be regarded as a logical continuation of the emerged research field. In contrast to the state-of-the-art work, we focus on learning the behavior at the application layer, which is subject to additional interdependencies on the different layers of the protocol stack.

%
%
\textbf{Anticipatory communication}: 
In previous work, we have explored network quality-aware channel access \cite{Ide/etal/2015a} and have demonstrated the massive potentials of using data rate prediction for optimizing the resource efficiency of delay-tolerant vehicular data transmissions \cite{Sliwa/etal/2018b, Sliwa/etal/2018a}.
%
%
Client-based data rate prediction within mobile cellular networks is a highly challenging task, as the resulting end-to-end throughput is influenced by various external and internal factors. In addition to mobility-related effects, which impact the channel coherence time, \emph{cross-layer} dependencies (e.g., the slow start mechanism of \ac{TCP}) have great influence on the observed end-to-end behavior \cite{Akselrod/etal/2017a}. 
\emph{Active} prediction methods monitor the data rates of ongoing data transmissions with time series-based analysis methods. As an example, \ac{TRUST} \cite{Wei/etal/2018a} brings together mobility pattern identification with \ac{TCP} data rate prediction based on \ac{LSTM} methods.
In contrast to that, \emph{passive} approaches only rely on measurable network quality indicators without introducing additional traffic themselves. In this paper, we focus on the passive measurement technique due to its wider acceptance within the research community, its better resource efficiency and its inherent capability of making predictions in an-hoc manner.
%
%
The authors of \cite{Jomrich/etal/2018a} analyze online data rate prediction based on a large data set for two different \acp{MNO} in a highway scenario.
%
%
Similar to Samba et al. \cite{Samba/etal/2017a}, the highest prediction accuracy is achieved with a \ac{RF} regression model. However, the resulting prediction accuracy is relatively low, as the end-to-end prediction is solely based on network context indicators and does not consider features, which are related to the cross-layer dependencies within the protocol stack of the \ac{UE}.
Similar studies are carried out by the authors of \cite{Riihijarvi/Mahonen/2018a}, which compare the performance of the machine learning models \ac{ANN}, \ac{LR}, \ac{GPR}, and \ac{RF}. Their findings conclude that these classic machine learning models -- with \ac{GPR} and \ac{RF} achieving the highest accuracies -- yield excellent prediction results, which can be utilized by the \mno to optimize its network processes.

%
%
\textbf{Maintaining network quality data}:
While the mobile \ac{UE} is able to perform measurements of the network quality indicators at its current location itself, it has to rely on estimation methods for forecasting those indicators at future locations. For this purpose, \emph{connectivity maps} \cite{Kelch/etal/2013a,Poegel/Wolf/2015a} can serve as a way for providing a data-driven method for maintaining geospatially-aggregated network quality information. 
%
%
In \cite{Sliwa/etal/2018a}, connectivity maps are jointly used with mobility prediction in order to schedule the time of vehicular sensor data transmissions with respect to the expected network quality on the future route.
%
%
Although it is possible to use and maintain these data bases in a completely decentralized way -- as people often drive the same routes regularly -- data freshness and the grade of covered areas can be significantly increased through exploitation of \emph{crowdsensing} approaches \cite{Wang/etal/2016a}. In order to increase the overall knowledge data base through using potentially heterogeneous data from different sources, correlation-based feature mapping \cite{Apajalahti/etal/2018a} can be applied. As an alternative to purely measurement-based approaches, the acquired data can be exploited to optimize the parameterization of radio propagation models. The latter are then exploited to estimate the network quality at unobserved locations. In \cite{Enami/etal/2018a}, Enami et al. present \ac{RAIK} as a method to forecast the \ac{RSRP}, which exploits highly detailed \ac{LIDAR} environment maps for achieving highly accurate estimations.

	\section{Methodology} \label{sec:methods}

In this section, the general \ac{DDNS} approach is introduced and the methodological aspects of the performance evaluation of the proposed method are described.

%
%
\fig{}{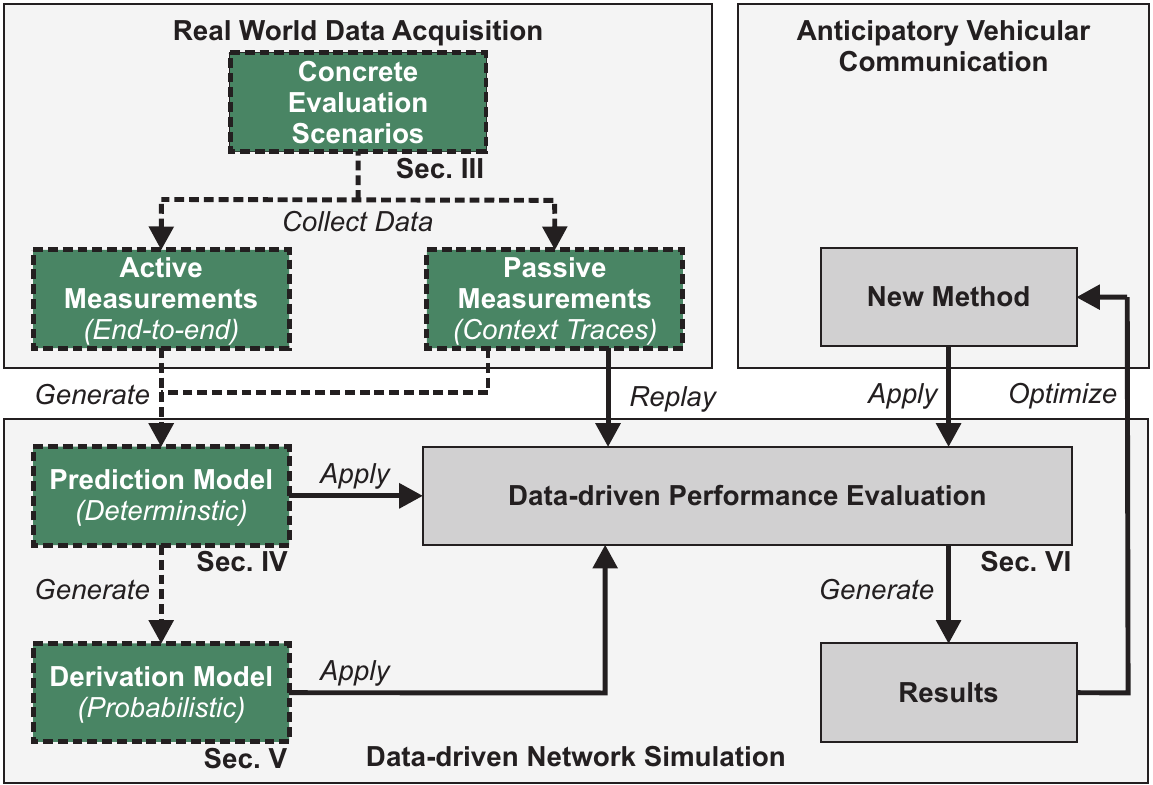}{Overall system architecture model and information flow for data-driven performance analysis and optimization. The dashed components are generated only once during the initial setup phase and are reused in the following steps.}{fig:ddns_architecture}

\subsection{Problem Definition and High-level Approach Description}

%
%
The overall goal of the proposed data-driven approach is to \emph{mimic} the network behavior of a \emph{concrete real world scenario}. For this purpose, \ac{DDNS} relies on \emph{replaying} previously acquired context traces (e.g., the measured network context indicators a vehicle has encountered on its trajectory) which are utilized to analyze the end-to-end performance of a \emph{novel} anticipatory communication method based on machine learning.

The logical information flow is illustrated in the overall system architecture model in Fig.~\ref{fig:ddns_architecture}.
%
%
\begin{itemize}
		
	%
	%
	\item \textbf{Prediction model generation:} In contrast to system-level network simulations which model actual communicating \emph{entities} including their protocol stacks, the proposed \ddns method relies on machine learning-based analysis of the end-to-end behavior. Supervised learning is applied to derive a \emph{deterministic} prediction model which allows to forecast the behavior of the considered end-to-end indicator based on the provided context traces. 
	In this work, we focus on data rate prediction in vehicular \ac{LTE} networks. Since the resulting accuracy of the prediction model is crucial for the achievable simulation accuracy, this aspect is analyzed detailedly in Sec.~\ref{sec:prediction}.

	%
	%
	\item \textbf{Derivation model generation:} If a data rate prediction model is applied in the real world, the actually achieved \emph{measurement} provides an immediately accessible ground truth for assessing the prediction accuracy. 
	As the defined goal of the \ac{DDNS} approach is to mimic the behavior of the real world network, the \emph{model imperfections} need to be taken into account within the simulations. However, since replaying the passive context traces implies to perform data rate prediction on \emph{unlabeled data}, a ground truth is missing. 
	For addressing this issue, a \emph{virtual measurement} is derived within the \ac{DDNS} by sampling from the error distribution of the real world measurements. For this purpose, a second machine learning model is applied to transform the prediction model from the deterministic to the \emph{probabilistic} domain. This process is further described in Sec.~\ref{sec:DDNS}.

	%
	%
	\item \textbf{Performance evaluation:} Finally, the performance evaluation is performed by applying the novel method on the replayed passive context measurements. The resulting end-to-end behavior is simulated based on the generated machine learning models. Sec.~\ref{sec:validation} illustrates the proposed methodological approach considering a case study focusing on opportunistic data transmission in vehicular networks.
\end{itemize}

%
%
\fig{b}{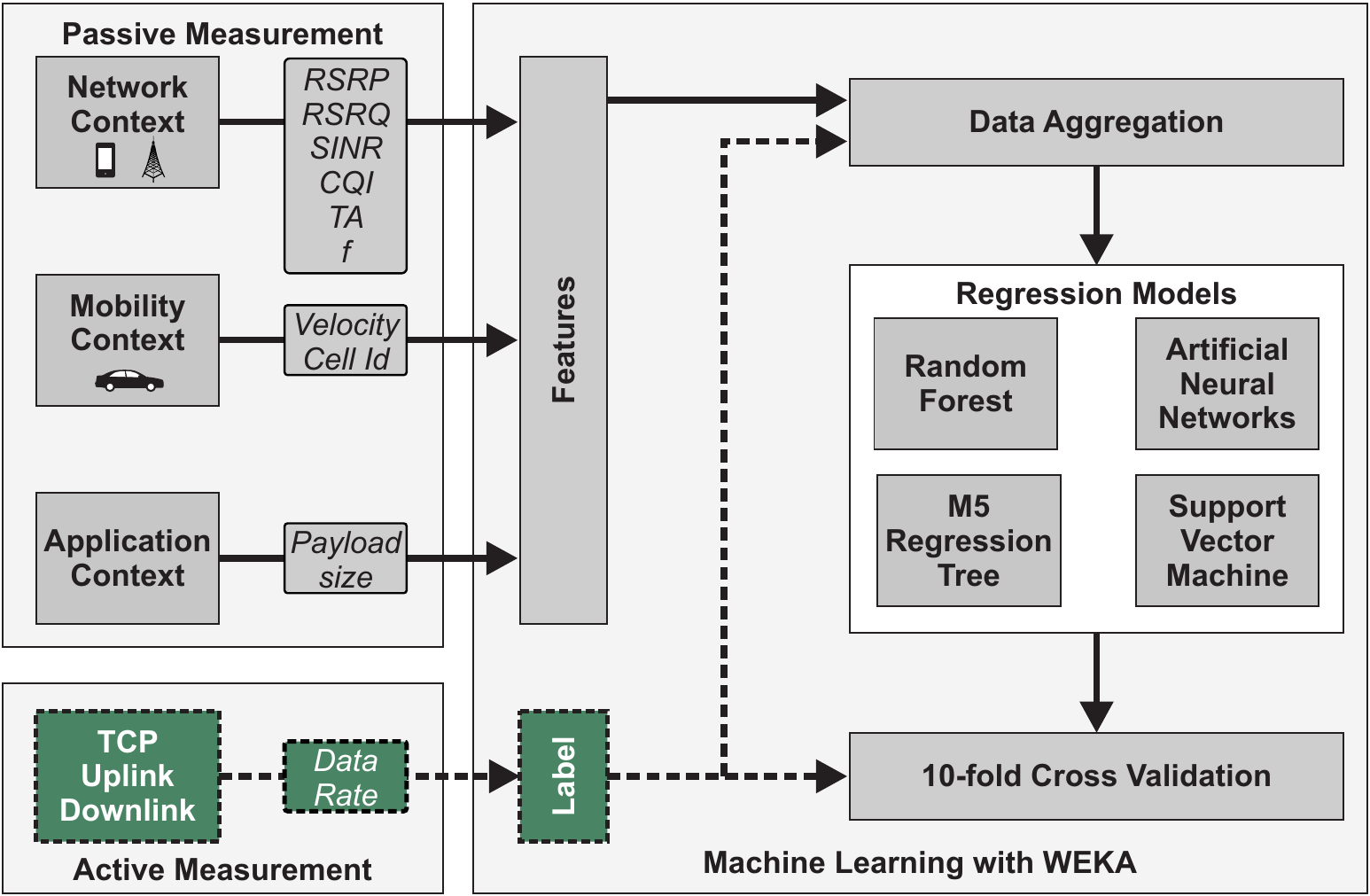}{Architecture model for the client-based data rate prediction.}{fig:prediction_architecture}

\subsection{Data Acquisition} \label{sec:data_acquisition}
For the later training of the machine learning models, a comprehensive data set is obtained by performing real world measurements in the public \ac{LTE} network of the three German \acp{MNO}. During the drive tests, every 10~s, a \ac{TCP}-based data transmission is performed with a random payload size in the range of ${0.1, 0.5, 1..10}$~MB in the uplink and in the downlink transmission direction. Furthermore, passive measurements of network quality indicators are acquired continuously.
The data rate measurement is handled at a remote server. All raw measurements can be accessed via \cite{Sliwa/2019a}.
The data transmissions are performed using multiple Android-based \acp{UE} (Samsung Galaxy S5 Neo, Model SM-G903F), which execute the developed measurement application\footnote{Measurement software available at https://github.com/BenSliwa/DDS}.
%
%
The real world drive tests are carried out in multiple scenarios, which differ with respect to the velocity range and the building density: \emph{campus} (3~km), \emph{urban} (3~km), \emph{suburban} (9~km), and \emph{highway} (14~km). Each track is driven ten times. In total, 12938 transmissions (58.45~GB of transmitted data) are performed on a total driven distance of 287~km.

\subsection{Data Analysis}
%
%
The machine learning-based data analysis is carried out with \ac{WEKA} \cite{Hall/etal/2009a} and \texttt{LIBSVM} \cite{Chang/Lin/2011a}. In order to automatically generate online prediction models as \texttt{C++} code from the abstract \ac{WEKA} results, we created a dedicated interface application, which is part of the supplied software package. If not stated otherwise, all presented data analysis results are 10-fold cross validated.

	\section{Client-based Data Rate Prediction} \label{sec:prediction}

\newcommand{\cW}{0.8cm}

\aboverulesep = 1pt
\belowrulesep  = 1pt

\begin{table*}[]
	\small
	
	\centering
	\caption{Coefficient of determination ($R^2$) for different machine learning models and data aggregation granularities.}
	\begin{tabular}{p{0.2cm}p{1.3cm}|p{\cW}p{\cW}p{\cW}p{\cW}|p{\cW}p{\cW}p{\cW}p{\cW}|p{\cW}p{\cW}p{\cW}p{\cW}p{\cW}}

		\toprule
		& & \multicolumn{4}{c|}{\textbf{\mno A}} & \multicolumn{4}{c|}{\textbf{\mno B}} & \multicolumn{4}{c}{\textbf{\mno C}} \\

		& \textbf{Data} & \textbf{ANN} & \textbf{M5} &\textbf{RF} & \textbf{SVM} & \textbf{ANN} & \textbf{M5} &\textbf{RF} & \textbf{SVM} & \textbf{ANN} & \textbf{M5} &\textbf{RF} & \textbf{SVM} \\
		\midrule	
		\sideHeader{5}{0.5cm}{Uplink} 
		& \textbf{\mno} & 0.685 & 0.754 & \textbf{0.8} & 0.71 & 0.46 & 0.658 & \textbf{0.707} & 0.594 & 0.69 & 0.779 & \textbf{0.82} & 0.728 \\
		& \textbf{Scenario} & 0.729 & 0.779 & \textbf{0.806} & 0.683 & 0.49 & 0.572 & \textbf{0.633} & 0.555 & 0.489 & 0.64 & \textbf{0.686} & 0.572 \\
		& \textbf{eNB} & 0.578 & 0.724 & \textbf{0.731} & 0.592 & 0.285 & 0.432 & \textbf{0.456} & 0.44 & 0.384 & 0.57 & \textbf{0.604} & 0.512 \\
		& \textbf{Cell} & 0.532 & 0.687 & \textbf{0.715} & 0.58 & 0.275 & 0.412 & \textbf{0.444} & 0.397 & 0.355 & \textbf{0.505} & \textbf{0.505} & 0.424 \\

		\midrule
		\sideHeader{6}{0.5cm}{Downlink} 
		& \textbf{\mno} & 0.499 & 0.603 & 0.591 & \textbf{0.612} & 0.524 & 0.584 & \textbf{0.648} & 0.578 & 0.41 & 0.504 & \textbf{0.552} & 0.531 \\
		& \textbf{Scenario} & 0.551 & 0.62 & 0.615 & \textbf{0.627} & 0.321 & 0.491 & \textbf{0.541} & 0.496 & 0.265 & 0.386 & \textbf{0.422} & 0.41 \\
		& \textbf{eNB} & 0.34 & 0.551 & 0.552 & \textbf{0.58} & 0.263 & 0.317 & 0.357 & \textbf{0.362} & 0.151 & 0.323 & 0.334 & \textbf{0.361} \\
		& \textbf{Cell} & 0.3 & \textbf{0.564} & 0.503 & 0.555 & 0.258 & 0.325 & \textbf{0.379} & 0.372 & 0.19 & 0.296 & \textbf{0.306} & 0.294 \\
	    \bottomrule		
		
	\end{tabular}
	
	\vspace{0.1cm}
	\emph{ANN}: Artificial Neural Network, \emph{M5}: M5 Regression Tree, \emph{RF}: Random Forest, \emph{SVM}: Support Vector Machine 
	\label{tab:throughput}
\end{table*}

This section discusses the prediction of the end-to-end data rate in uplink and downlink direction in multi-\mno networks. The availability of reliable prediction models is one of the foundations of the proposed \ac{DDNS} approach, which is further discussed in Sec.~\ref{sec:DDNS}.

%
%
Predicting end-to-end performance indicators is a \emph{regression} task, where a model $f$ is trained to learn the relationship between a \emph{feature} set $\mathbf{X}$ and a \emph{labeled} data set $\mathbf{Y}$. After the training phase, the model can be utilized to make predictions $\tilde{y}$ on new data $\mathbf{x}$ such that $\tilde{y} = f(\mathbf{x})$.

The overall architecture model of the machine learning-based data rate prediction process, which is conducted in this paper, is illustrated in Fig.~\ref{fig:prediction_architecture}. 
In the following evaluations, the feature set $\mathbf{X}$ is composed of nine features from different logical context domains:
%
%
\begin{itemize}
	\item The \textbf{application context} consists of the payload size of the data packets, which are transmitted via \ac{TCP}.
	\item The \textbf{channel context} is formed by the passive \ac{LTE} network quality indicators \ac{RSRP}, \ac{RSRQ}, \ac{SINR}, \ac{CQI}, \ac{TA} and the carrier frequency of the serving \ac{eNB}.
	\item The \textbf{mobility context} is represented by the vehicle's velocity and the current cell id.
\end{itemize}
%
%
During the training phase, the resulting data rate of the active transmissions is utilized as the labeled data set $\mathbf{Y}$. The actual regression task is performed by multiple machine learning models, which were tuned in a preparatory step.
%
%
\begin{itemize}
	\item \textbf{\acf{ANN}} \cite{LeCun/etal/2015a}, where a deep neural network with two hidden layers (10 and 5 neurons) showed the highest prediction accuracy. Learning rate $\eta=0.1$ and momentum $\alpha=0.001$ were optimized based on an evolutionary algorithm.
	\item \ac{CART}-based models: \textbf{\acf{RF}} \cite{Breiman/2001a}, which consists of 100 random trees of maximum depth 20 and \textbf{\acf{M5}} \cite{Quinlan/1992a}. 
	\item \textbf{\acf{SVM} with \acf{RBF} kernel} \cite{Cortes/Vapnik/1995a} trained with \ac{SMO} regression.
\end{itemize}
%
%
%

%
%
For completeness, it is remarked that other regression models such as \ac{KNN} and \ac{LR} were also considered during the initial model exploration phase. However, as those approaches did not reach a performance level comparable to the other -- and more widely used -- data rate prediction models, they were excluded from the deeper evaluations. The interested reader is forwarded to \cite{Nikolov/etal/2018a, Riihijarvi/Mahonen/2018a}

%
%
As a statistical metric for the model performance and for allowing a comparison to related work  (e.g., \cite{Samba/etal/2017a, Jomrich/etal/2018a}), which consider the same performance indicator, the \emph{coefficient of determination} is analyzed. It is calculated as 
%
%
\begin{eqnarray}
	R^{2} = 1- \frac{\sum_{i=1}^{N}\left(\tilde{y}_{i} - y_{i} \right)^{2}}{\sum_{i=1}^{N}\left(\bar{y} - y_{i} \right)^{2}}
\end{eqnarray}
with $\tilde{y}_{i}$ being the current prediction, $\bar{y}$ the mean of the measurement and $y_{i}$ the current measurement. The $R^{2}$ describes the amount of the response variable variation, which is explained by the derived regression model.

%
%
\newcommand{\sfw}{0.32}
\begin{figure*}[] 
	\centering
	
	%
	%
	\subfig{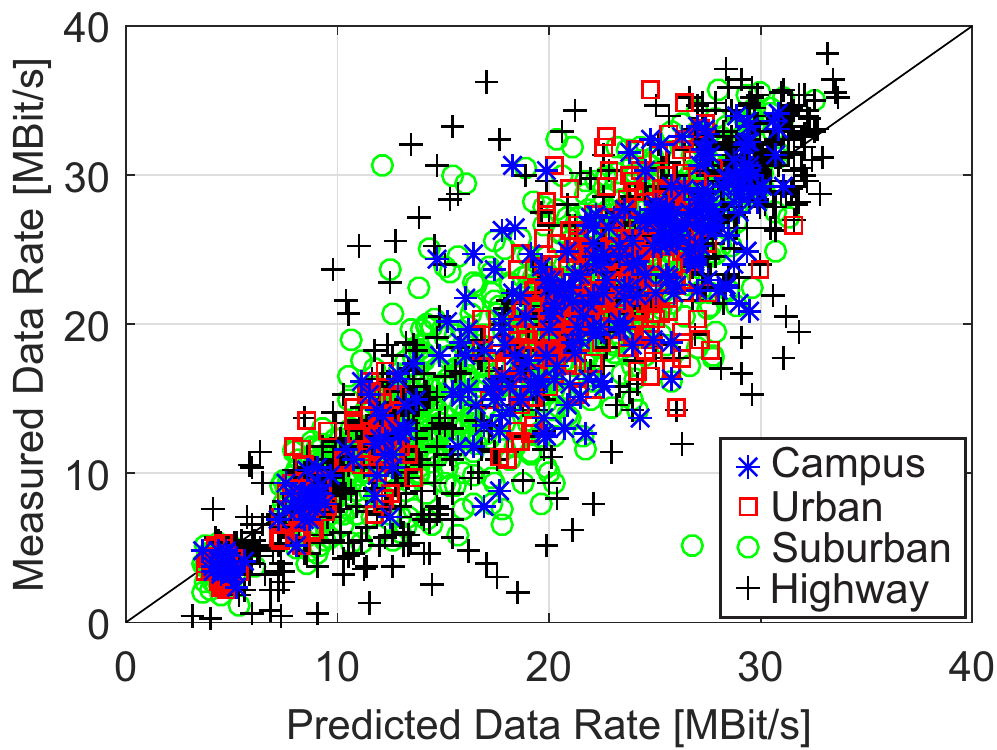}{\sfw}{MNO~A (Uplink)}
	\subfig{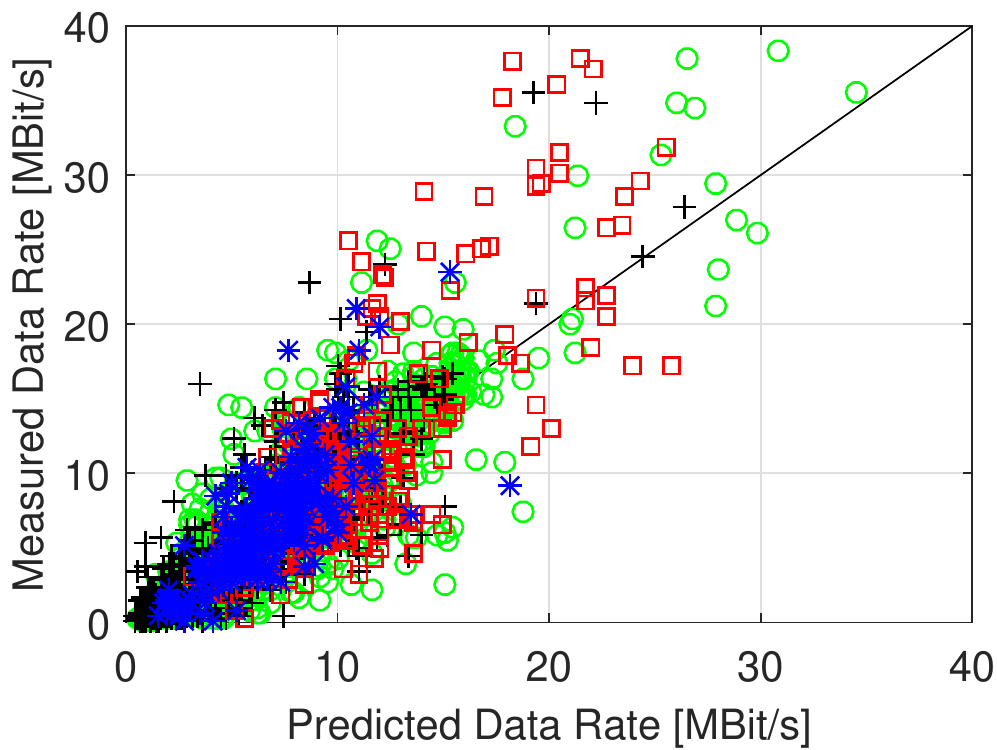}{\sfw}{MNO~B (Uplink)}
	\subfig{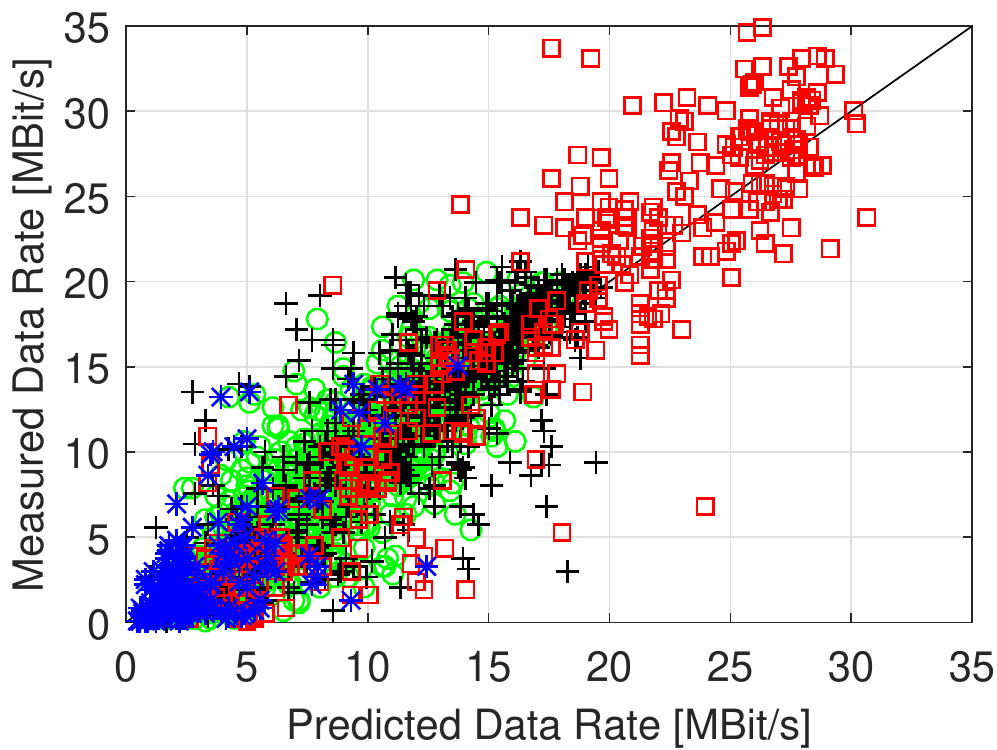}{\sfw}{MNO~C (Uplink)}

	%
	%
	\subfig{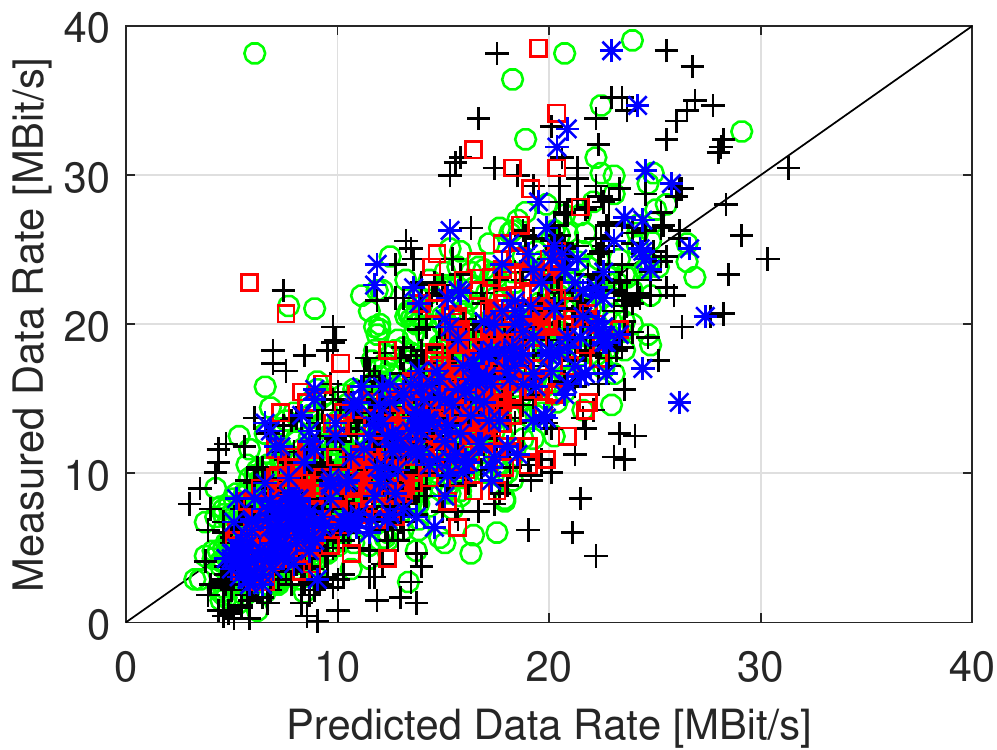}{\sfw}{MNO~A (Downlink)}
	\subfig{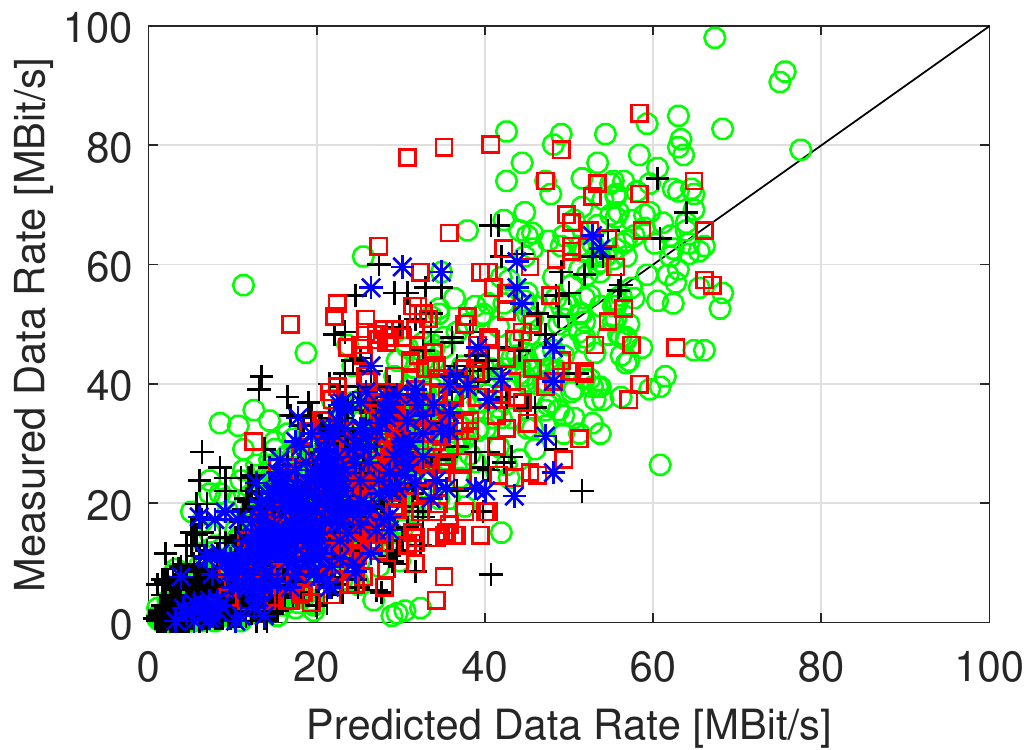}{\sfw}{MNO~B (Downlink)}
	\subfig{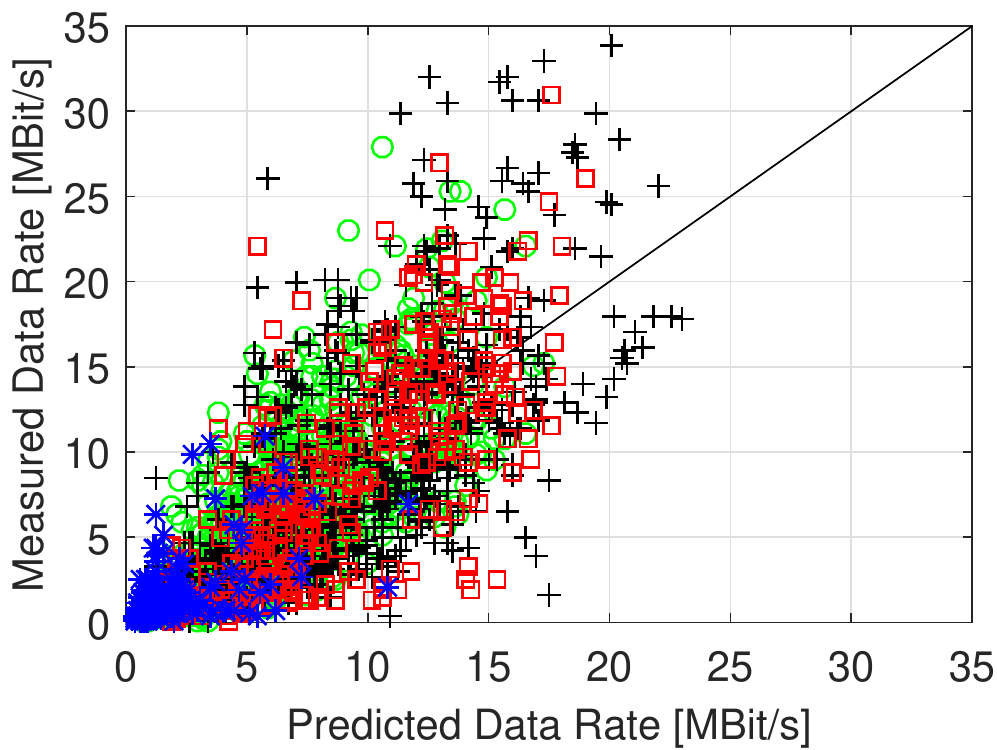}{\sfw}{MNO~C (Downlink)}
	
	\caption{Measured transmission profiles for \ac{RF}-based data rate prediction in uplink and downlink direction in different evaluation scenarios.}

	\label{fig:model_performance}
\end{figure*}

\subsection{Comparison of Different Prediction Models and Training Data Granularities}

In the first evaluation, the overall training data set is split into various subsets in order to find the most usable data aggregation granularity within the trade-off between using a higher amount of training data -- e.g., a single global data set per \mno -- or focusing deeper on the infrastructure-specific aspects, which would imply to utilize many local data sets. In addition, it is analyzed, which regression model achieves the highest prediction accuracy and will be utilized in the further evaluation phases.

For both transmission directions, all regression models are trained on all data subsets, which are composed as follows:
%
%
\begin{itemize}
	\item \textbf{\mno} (3 sets): Global data set per \mno
	\item \textbf{Scenario} (12 sets): Evaluation track-specific data aggregation (campus, urban, suburban, highway)
	\item \textbf{\enb} (105 sets): Data aggregation based on the \enb id
	\item \textbf{Cell} (220 sets): Data grouping based on the cell id
\end{itemize}
Tab.~\ref{tab:throughput} summarizes the $R^2$ results of the resulting prediction performance for all variants.

%
Overall, it can be seen that the highest prediction accuracy is achieved with the \ac{CART}-based models \ac{RF} and \ac{M5}, which is confirmed by the findings of related performance evaluations \cite{Samba/etal/2017a, Jomrich/etal/2018a}. 
As pointed out in the analysis of \cite{Sliwa/etal/2018b}, in many cases, a single network quality indicator has a dominant impact on the resulting data rate under well-defined conditions. While the \ac{SINR} is an important indicator within the cell center region, the \ac{RSRQ} has a major impact on the considered end-to-end indicator at the cell edge. The regions themselves can be estimated with the \ac{RSRP} which is depending on the distance to the serving \ac{eNB}.
Since the \ac{CART} models provide a scope-wise feature hierarchy within their model structure, they are able to represent these conditions in their native model architecture.

%
%
In addition to the achieved accuracy, a great advantage of the \ac{CART}-based models is that they can be implemented in a highly resource efficient way using simple \texttt{if/else} statements. Within the online application of the trained models, the execution time for making predictions is nearly negligible. On the considered Android platform, the average online execution time per single prediction is $\sim0.1$~ms for the trained \ac{RF}. The training of the 10-fold cross validation is performed in less than a minute.
%
%
Although the \ac{RF} achieves the highest prediction accuracy, it is remarkable that the much simpler \ac{M5} is often only slightly less accurate. As an example for uplink prediction of \mnoA, the trained \ac{RF} consists of $120533$ leafs, which contain numerical values. The trained \ac{M5} only consists of 11 leafs, which contain linear regression models. The lightweight model size of the \ac{M5} can be exploited for enabling the usage of machine learning even on highly resource constrained systems (e.g., microcontrollers). 

%
%
As the analysis shows, in most cases, the considered regression models benefit more from using a higher amount of training data than from increasing the grade of locality. 
%
%
Based on the obtained results, the following evaluations focus on a deeper analysis of the \ac{RF} regression model with the global data sets for each \mno and transmission direction.

\subsection{Behavior Analysis of the Random Forest Data Rate Prediction Model} \label{sec:rf_behavior}

%
\renewcommand{\sfw}{0.24}
\begin{figure*}[] 
	\centering
	
	\subfig{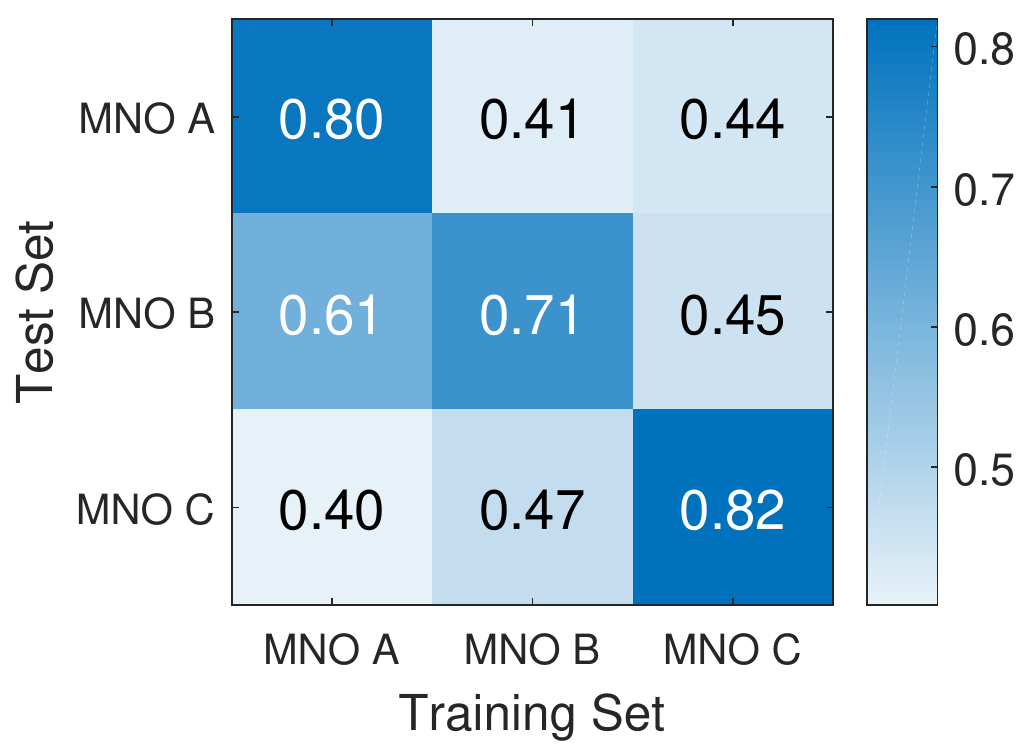}{\sfw}{Cross-\mno (Uplink)}%
	\subfig{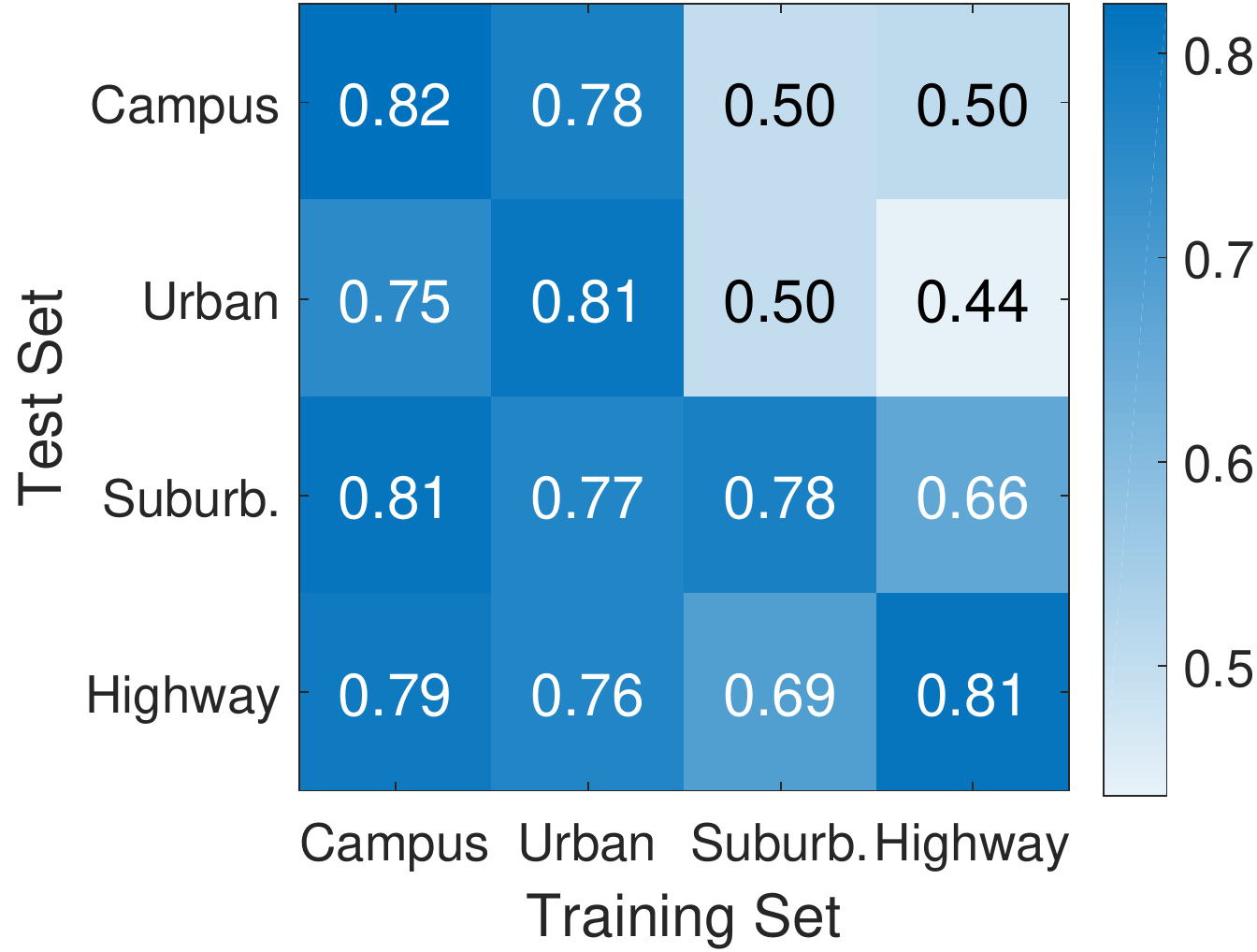}{\sfw}{\mno~A (Uplink)}%
	\subfig{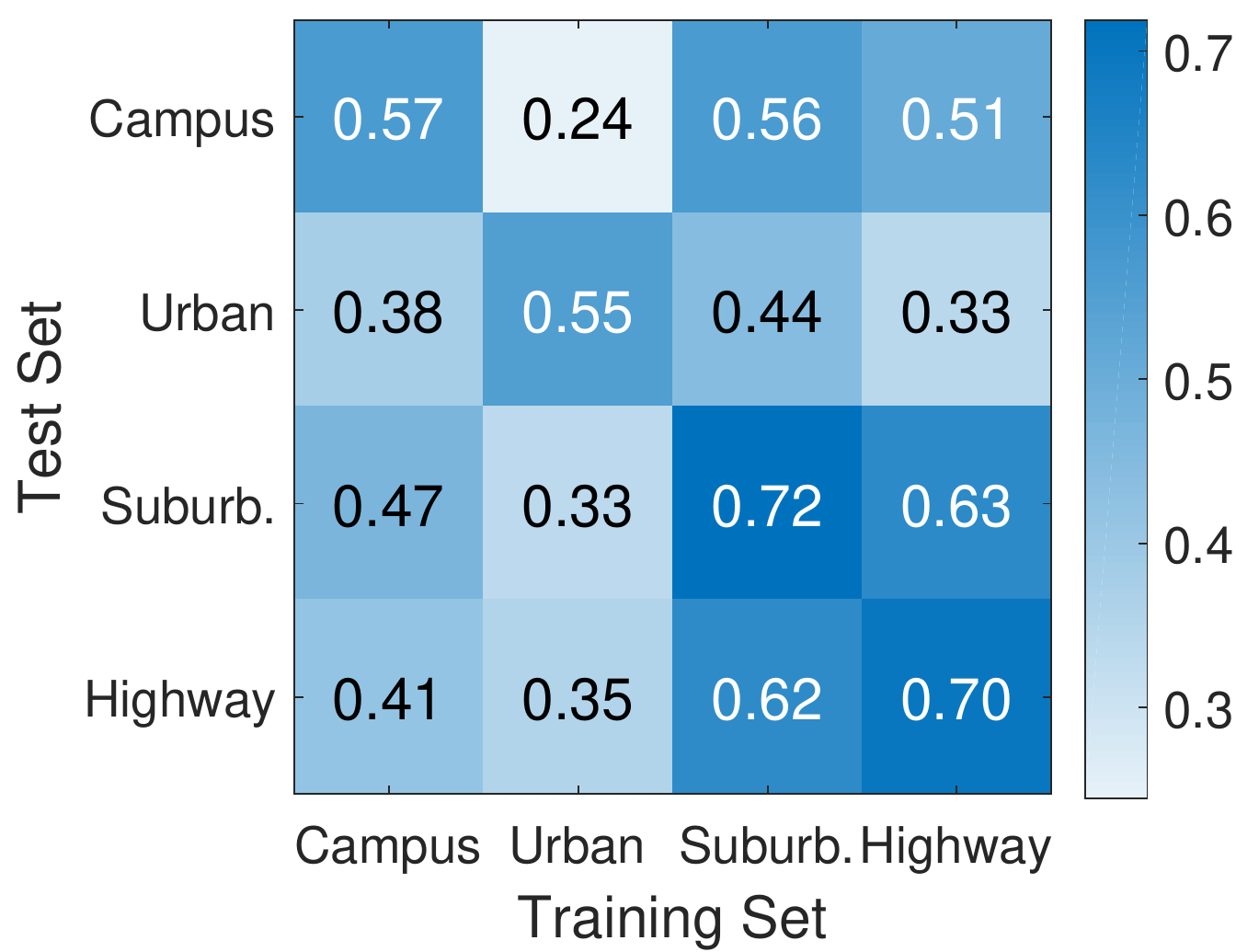}{\sfw}{\mno~B (Uplink)}%
	\subfig{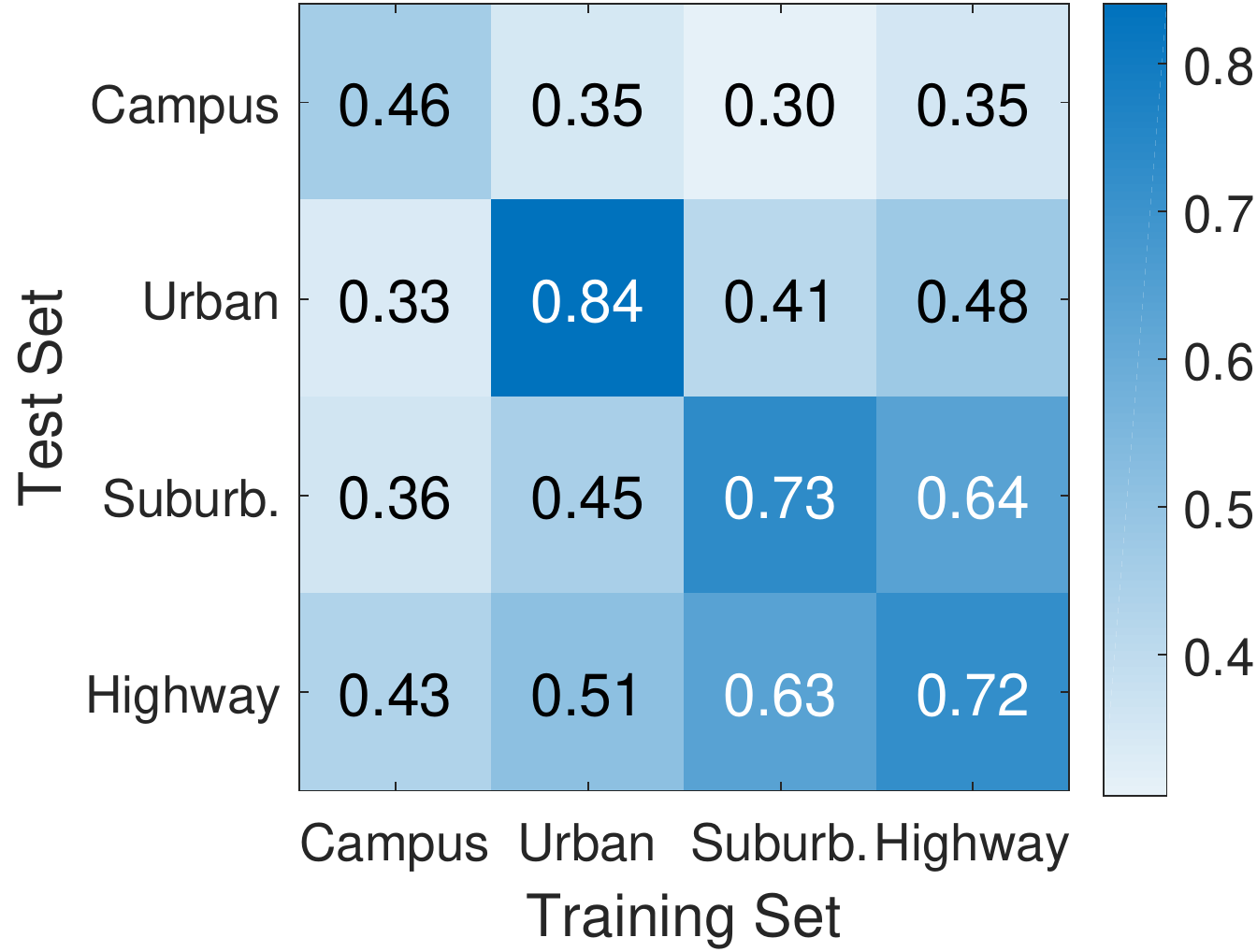}{\sfw}{\mno~C (Uplink)}%

	\subfig{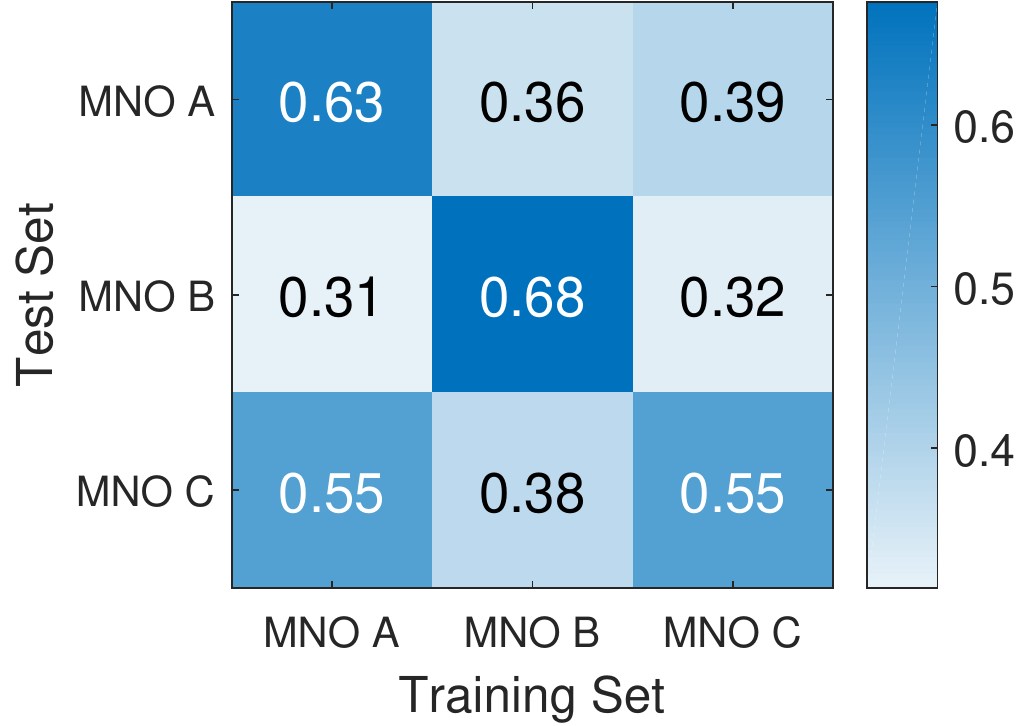}{\sfw}{Cross-\mno (Downlink)}%
	\subfig{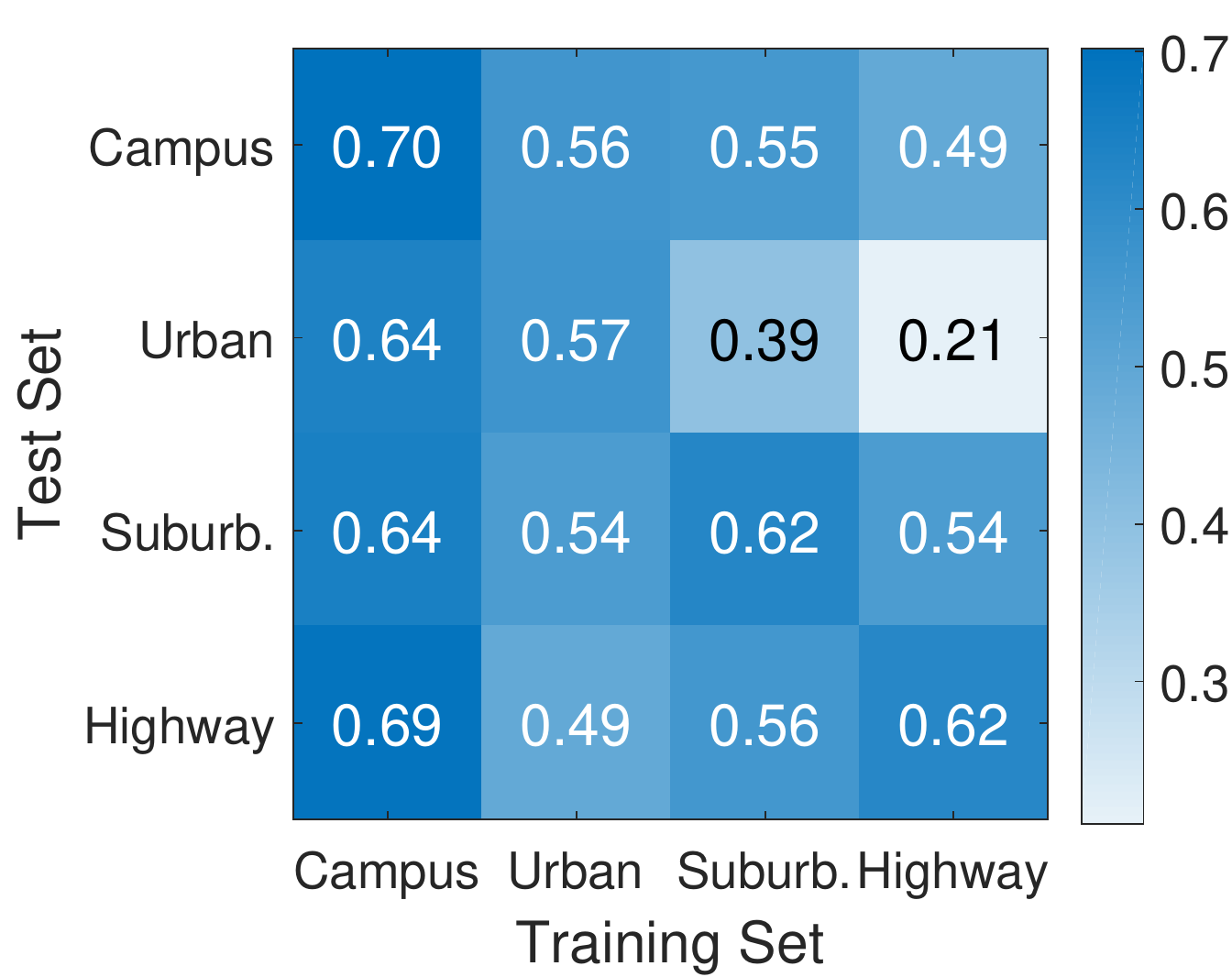}{\sfw}{\mno~A (Downlink)}%
	\subfig{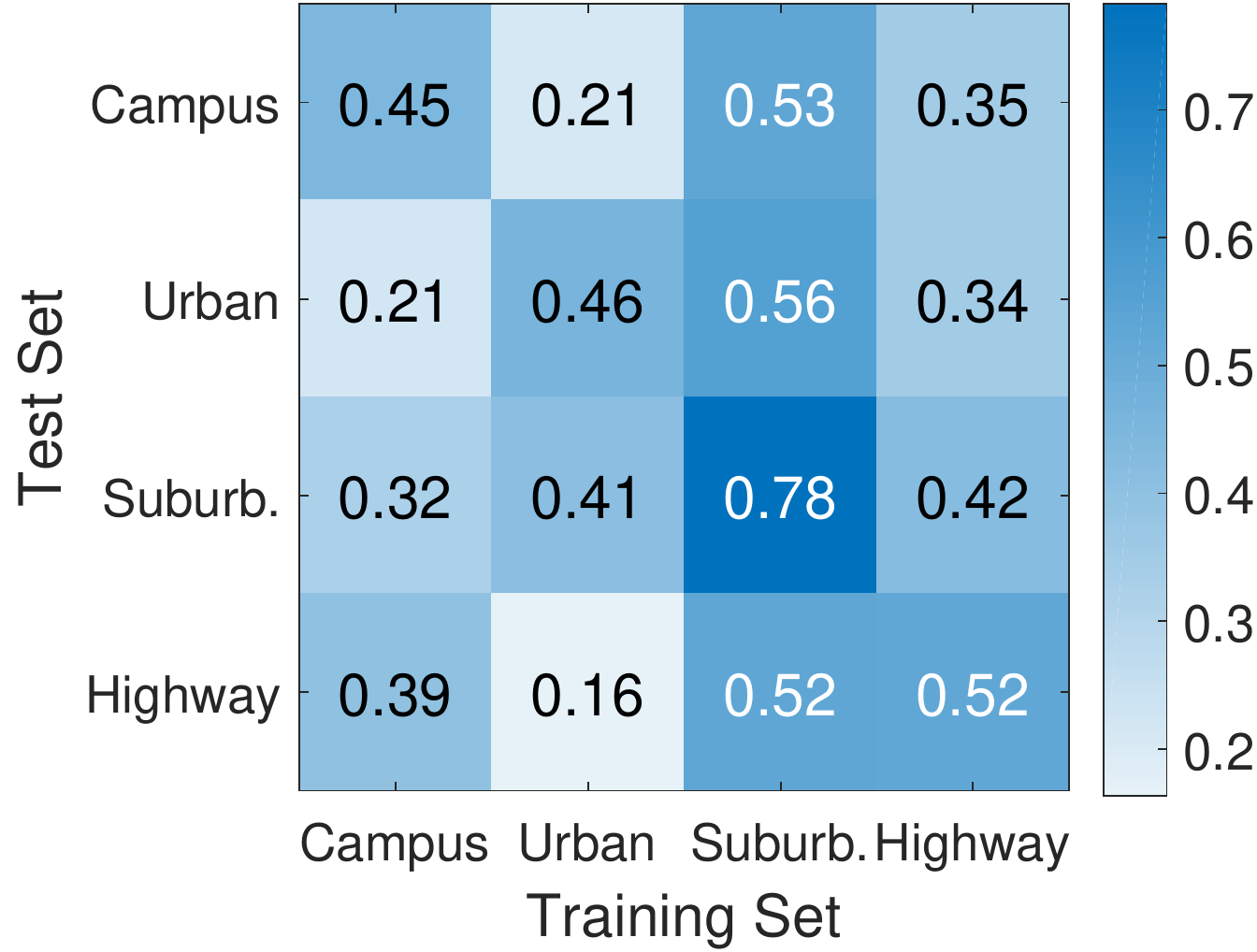}{\sfw}{\mno~B (Downlink)}%
	\subfig{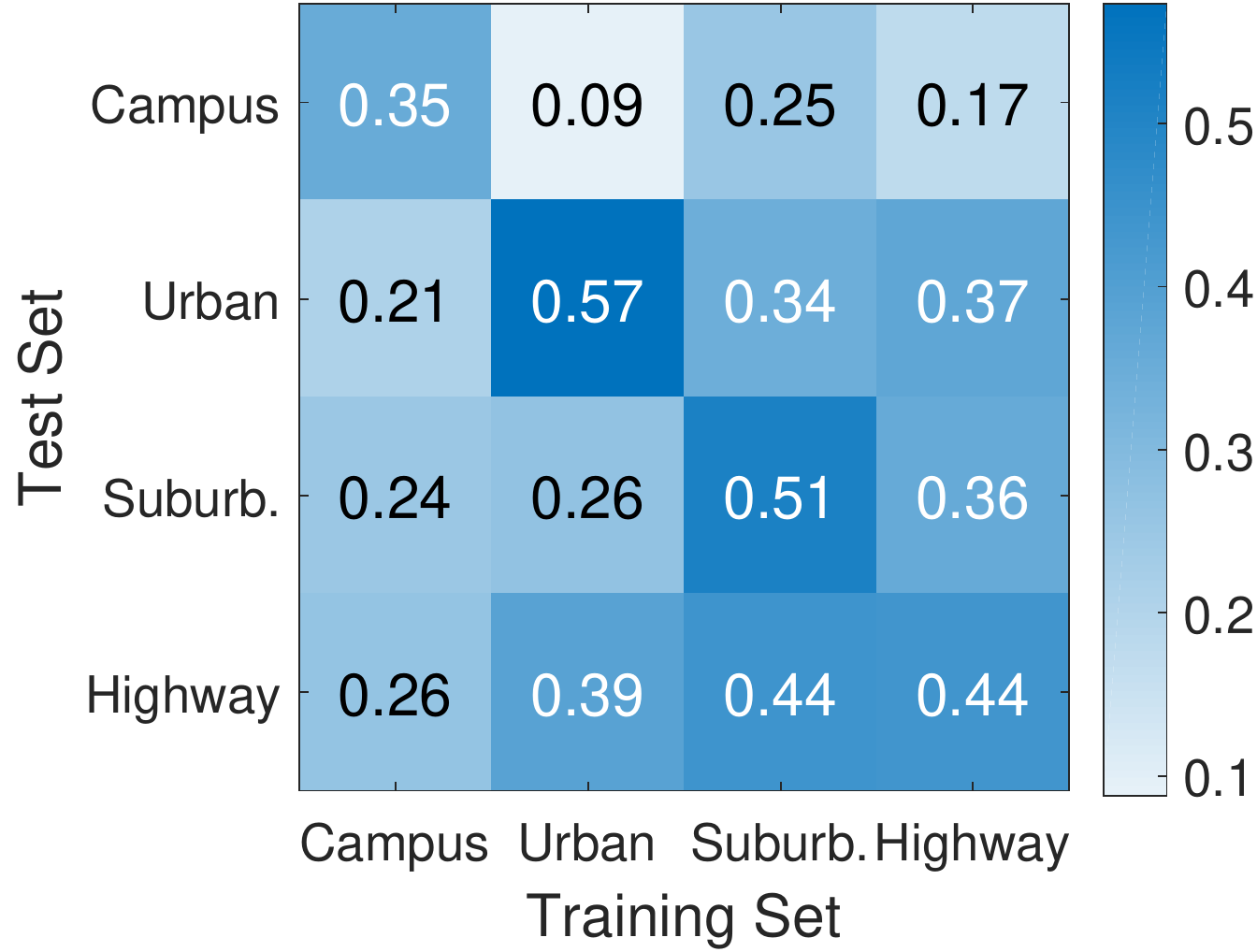}{\sfw}{\mno~C (Downlink)}%
	
	\caption{Coefficient of determination ($R^2$) results for the cross-\mno and cross-scenario prediction performance. The main diagonal elements show the 10-fold cross validation results, all other elements have distinct training and tests sets.}
	\label{fig:confusion}
\end{figure*}
The resulting prediction performance of the \ac{RF} models of each \mno in uplink and downlink direction is visualized in Fig.~\ref{fig:model_performance}. It can be seen that the behavior is highly depending on the \mno and its provided coverage within each scenario.
%
%
For \mnoA, the values are spread homogeneously for all scenarios. In contrast to that, \mnoB and \mnoC have focus regions, where a distinct level of performance is provided (e.g., \mnoC only provides the highest performance in the urban scenario).
Overall the highest spread of the prediction error can be observed in the highway scenario. Due to the high velocity range up to 150~km/h, the channel coherence time is low and handovers occur frequently.
%
%
Apart from \mnoA, which achieves a similar performance in both transmission directions, it can also be observed that the operators prioritize uplink and downlink performance differently. \mnoB is the only operator, which provides downlink \ac{CA}. Therefore, the value range of the downlink measurements is significantly larger than for the other \acp{MNO}.

%
%
\begin{figure}[]
	\centering		  
	
	\subfig{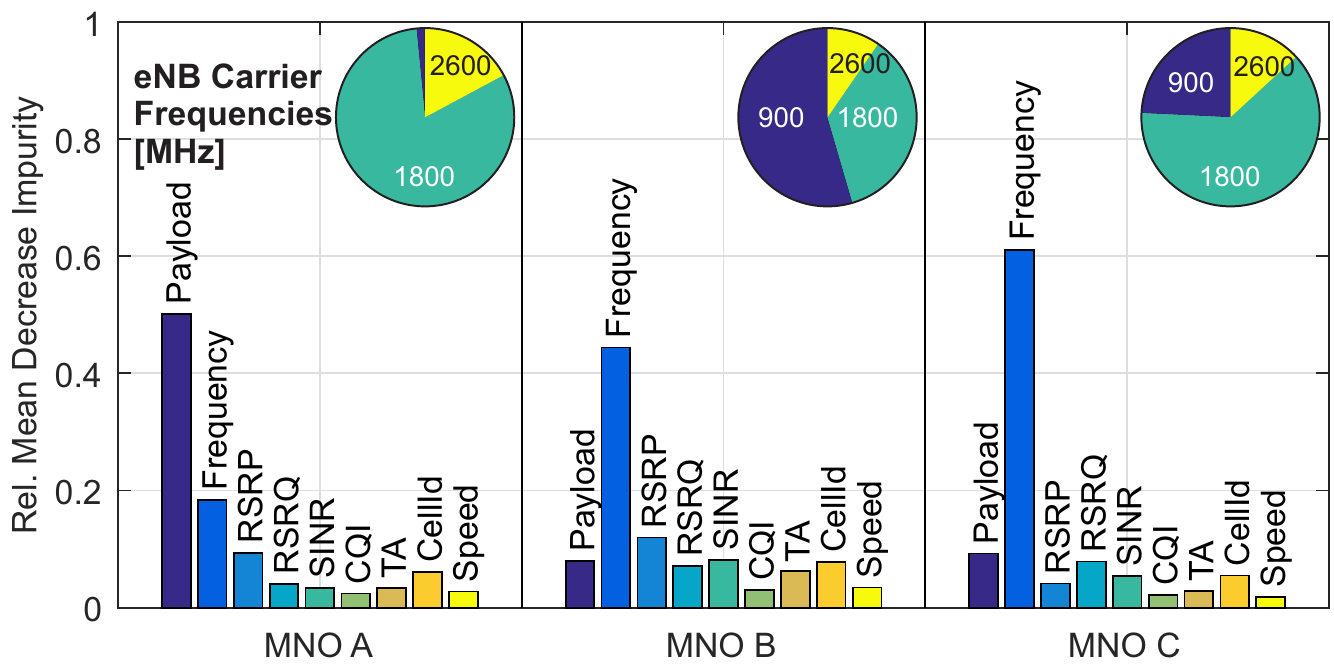}{0.48}{Uplink}%
	\subfig{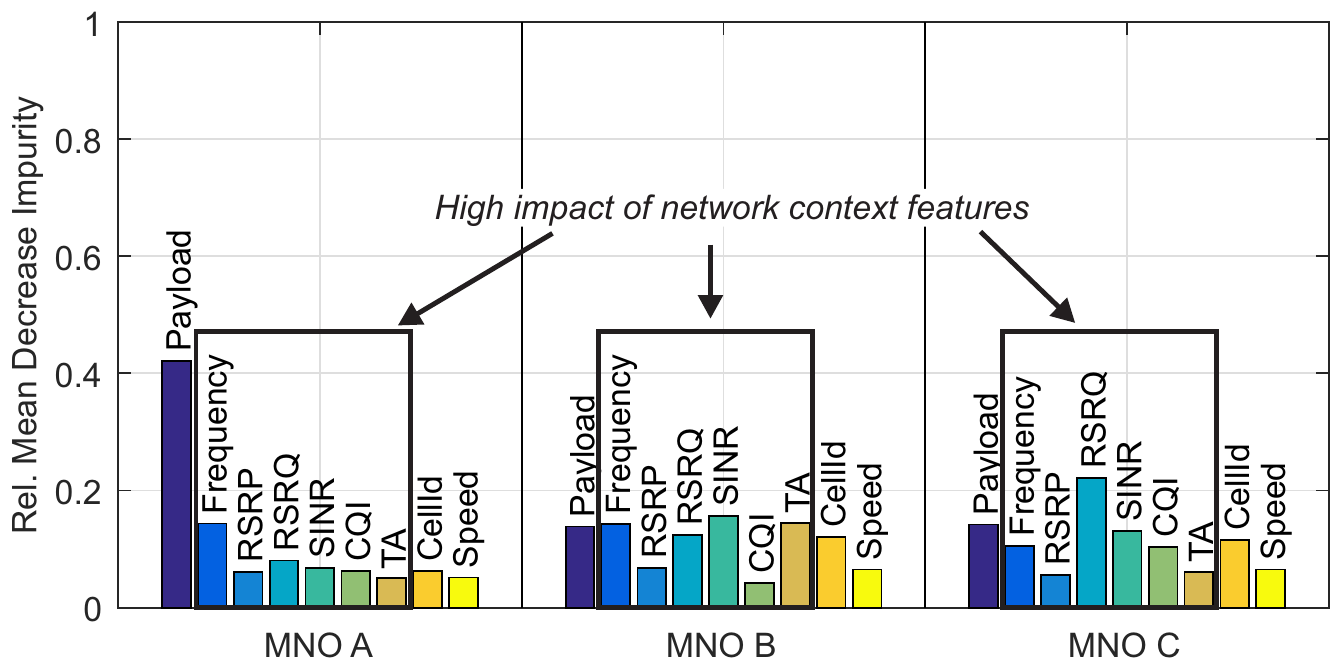}{0.48}{Downlink}%
	
	\caption{Importance of individual features for the overall prediction accuracy. The overlay shows the distribution of the \ac{eNB} carrier frequencies for each \mno.}
	\label{fig:featureImportance}
\end{figure}

%
%
\begin{figure*}[] 
	\centering
	\includegraphics[width=1\textwidth]{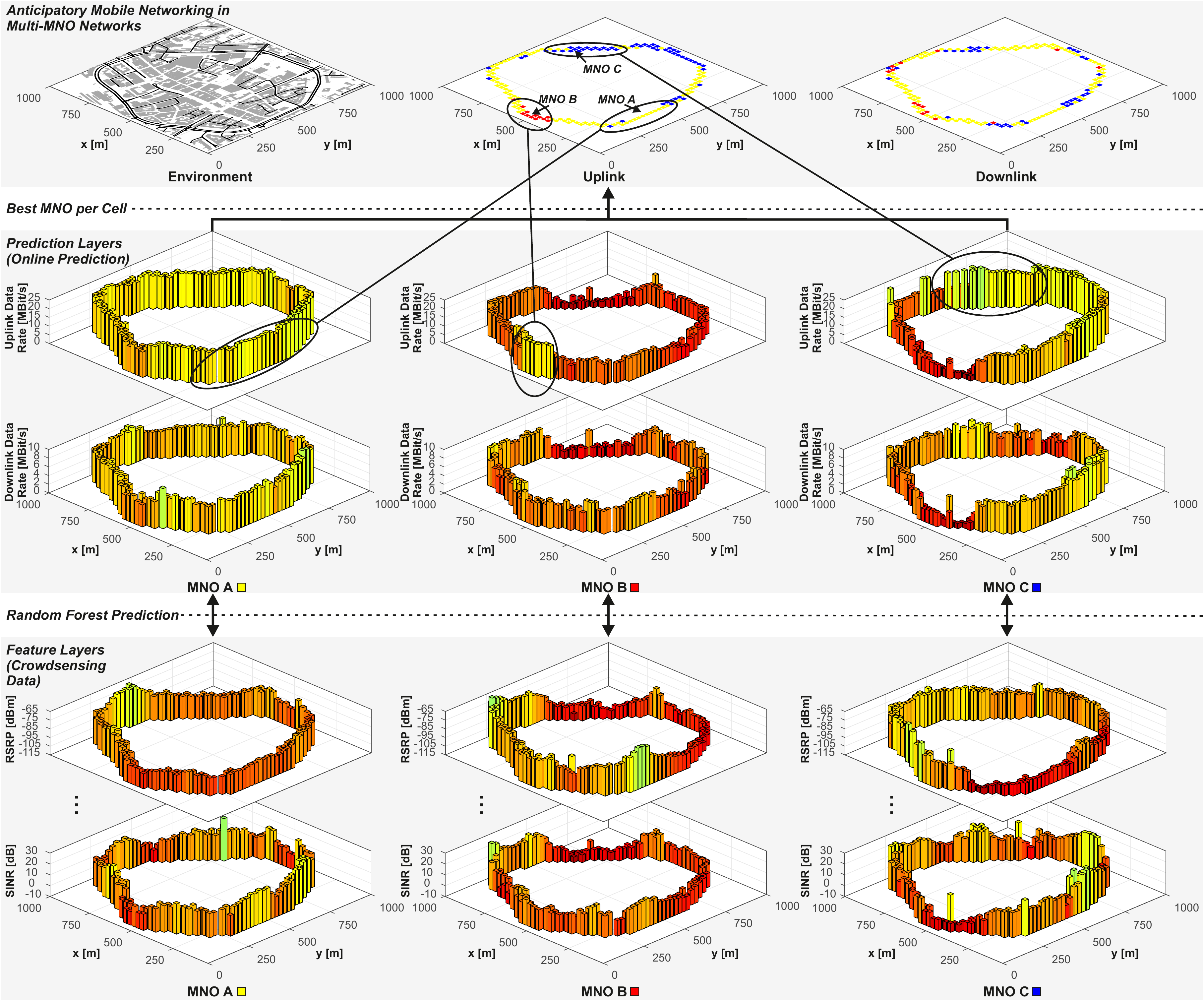}	
	\caption{Excerpt of the multi-\mno connectivity map for the urban scenario. For the data rate prediction a payload size of $2$~MB is assumed. Due to spacial limitations, the feature layers are only shown for the \ac{RSRP} and the \ac{SINR}. The actually applied connectivity map consists of nine different features. (Map data: \textcopyright OpenStreetMap contributors, CC BY-SA).}

	\label{fig:cm}
\end{figure*}
%
%
%

%
%
Fig.~\ref{fig:confusion}(a) and Fig.~\ref{fig:confusion}(e) show the resulting $R^2$ for cross-\mno data rate prediction in uplink and downlink transmission direction. It can be observed that the learned models are only able to provide significant results for the networks of the \mno they were trained on. It can be concluded that the measurable context indicators have to be considered jointly with the non-measurable \mno-specific configurations, which are implicitly learned as \emph{hidden features}.

%
%
Fig.~\ref{fig:confusion}(b)-(d) and Fig.~\ref{fig:confusion}(f)-(h) show the \ac{MNO}-specific cross-scenario prediction performance. For each \mno, a \ac{RF} model is trained on the data subset of each scenario and tested against the other scenarios. 
%
%
For \mnoA, the campus and urban subsets achieve very good generalization for all test sets. However, the data subsets for the highway and the suburban scenarios do not generalize well. Considering Fig.~\ref{fig:model_performance}(a) and Fig.~\ref{fig:model_performance}(d), it can be seen that the error spread is significantly higher for those two scenarios than for the others. Therefore, prediction artifacts, which arise from the low channel coherence time in the challenging environments, limit the cross-scenario prediction accuracy. In contrast to that, the other subsets succeed better on learning the general impact between context indicators and resulting data rate. In addition, the \ac{LTE} cells in the campus and urban subsets are more crowded than in the suburban and highway subsets. Therefore, if only the latter scenarios are considered, the machine learning model fails to learn the interdependency between cell load -- through measurements of the \ac{RSRQ} -- and data rate for high load scenarios within congested cells.
%
%
For \mnoB and \mnoC, the cross-scenario generalization is low, as the network performance itself is highly scenario-dependent (see Fig.~\ref{fig:model_performance}). Moreover, \ac{LTE} coverage is not always guaranteed, e.g., \mnoC suffers from poor \ac{LTE} coverage (76.25~\%) in the campus scenario.

%
%
The results emphasize that meaningful data sets should be composed of data from different heterogeneous scenarios in order to achieve good generalization. However, it is not reasonable to handle the different scenarios with scenario-specific prediction models. In all cases, the global \mno data sets achieve a higher mean $R^2$ than the overall average $R^2$ of all individual scenarios.

\subsection{Impact of Individual Features} \label{sec:featureImportance}

For assessing the impact of individual features on the resulting prediction accuracy, the relative \ac{MDI} \cite{Louppe/etal/2013a} is computed for the different \acp{RF}. The results of the evaluations are shown in Fig.~\ref{fig:featureImportance}.

%
%
%
It can be seen that the feature importance is depending on the \mno. It is influenced by the unknown resource scheduling policy and the unknown configurations of the hardware components of the network infrastructure itself. 
%
%
While the carrier frequency has a dominant impact on the uplink prediction accuracy for \mnoB and \mnoC, the feature is less important for \mnoA. As the overlayed distribution of the observed carrier frequencies shows, the \ac{UE} is mostly connected to 1800~MHz cells in the network of \mnoA. For the other \acp{MNO}, the carrier frequencies are distributed more diversely.
%
%
In the downlink direction, the importance of the carrier frequency is significantly reduced for \mnoB and \mnoC. While it is possible that the \acp{eNB} employ different scheduling policies for uplink and downlink, another explanation is the traffic pattern of the cell users. As the downlink resources are more often subject to resource competition \cite{Bui/etal/2017a}, it is plausible that the radio propagation-related impact is less significant than the resource allocation process. For \mnoA, the feature importance is symmetrical for uplink and downlink.

%
%
In comparison to related work \cite{Samba/etal/2017a, Jomrich/etal/2018a}, the achieved overall prediction accuracy is significantly higher. While the mentioned approaches only consider the network context features for the prediction, other dominant influences such as the payload size are not considered.
%
%
The achievable average data rate of a transmission is directly related to the payload size as the latter has a strong impact on the resulting transmission time and the behavior of the \ac{TCP} slow start mechanism. In the vehicular context, the \ac{UE} moves during the transmission process, which results in a low channel coherence time. While larger payload sizes are beneficial from a transport layer perspective \cite{Falkenberg/etal/2018a}, higher transmission durations increase the probability of significant changes of the channel quality during active transmissions. However, these complex cross-layer interdependencies are implicitly considered by the applied machine learning-based approach.

%
%
For completeness, it is remarked that the integration of additional features (e.g., time of day) was analyzed in a pre-evaluation step. As their consideration did not increase the resulting prediction accuracy, they were removed from the feature set. This behavior can be explained by their correlation to already contained features. As an example, the time of day can be used as an indicator for the load dynamics of the \ac{LTE} network \cite{Bui/etal/2017a}, but similar information is provided by the \ac{RSRQ}, which is already contained in the feature set.

%
%
Within upcoming 5G networks, the \ac{NWDAF} \cite{3GPP/2019a} of the core network will act as machine learning-based method for estimating the load level of network slices. Although similar analyses can already be performed by the \acp{UE} using passive control channel analysis \cite{Bui/Widmer/2016a}, providing the \ac{NWDAF} information itself for the cell users could greatly improve client-side data rate prediction and would therefore significantly contribute to catalyzing anticipatory mobile networking techniques.

\subsection{Exploiting Crowdsensing Data For Network Quality Prediction}

%
%
The presented prediction methods rely on immediate measurements of different context indicators, which allow to derive data rate predictions only for the current vehicle location. However, state-of-the-art anticipatory communication techniques are able to significantly benefit from exploiting knowledge about the network quality along the expected future trajectory (e.g., for opportunistic data transfer \cite{Sliwa/etal/2018a}, which is applied for the \ddns validation in Sec.~\ref{sec:mus_validation}). 

%
%
Since the vehicle itself is not able to measure the network quality at the future locations, it has to rely on previously obtained spatially aggregated data, which can be provided by crowdsensing-based connectivity maps. 
Fig.~\ref{fig:cm} shows an excerpt of the derived multi-\mno connectivity map for the urban evaluation track. The connectivity map is organized into three logical layers.
The lowest layer consists of previous measurements of the individual features of the prediction scheme. Each cell of the connectivity map contains the aggregated information of measurements, which were performed in the same cell during previous drive tests or by other network participants. For a defined cell size $c$ and a given position prediction $\mathbf{\tilde{P}}(t+\tau)$, the cell key $k$ is computed as
%
%
\begin{equation}
	k = \lfloor \frac{\mathbf{\tilde{P}}(t+\tau)}{c}\rfloor
\end{equation}
and utilized to access the context information $\mathbf{C}$ from the connectivity map.
%
%
The \emph{prediction layers} maintain the prediction results of the considered end-to-end indicators for each \mno and are based on the feature layer information.
On the highest layer, the prediction results are exploited by anticipatory networking techniques. In the considered example shown in Fig.~\ref{fig:cm}, the availability of multiple \acp{MNO} is exploited for data rate-aware interface selection.

%
%
Apart from enabling context-predictive networking methods, the usage of connectivity maps for maintaining the feature information has additional advantages. First, it allows to separate measurement platform and application platform. Although not all \ac{UE} types and operating systems are able to provide the same network quality indicators \cite{Falkenberg/etal/2018a}, anticipatory networking methods can still exploit this information if it has been measured by other \acp{UE} and is maintained by a connectivity map. Second, it enables the usage of synthetic mobility traces \cite{Malandrino/etal/2018a} for evaluating the to be analyzed method at unobserved locations.

	\section{Data-driven Simulation of End-to-end Network Performance Indicators} \label{sec:DDNS}

%
%
\begin{figure}[]
	\centering		  
	
	\subfig{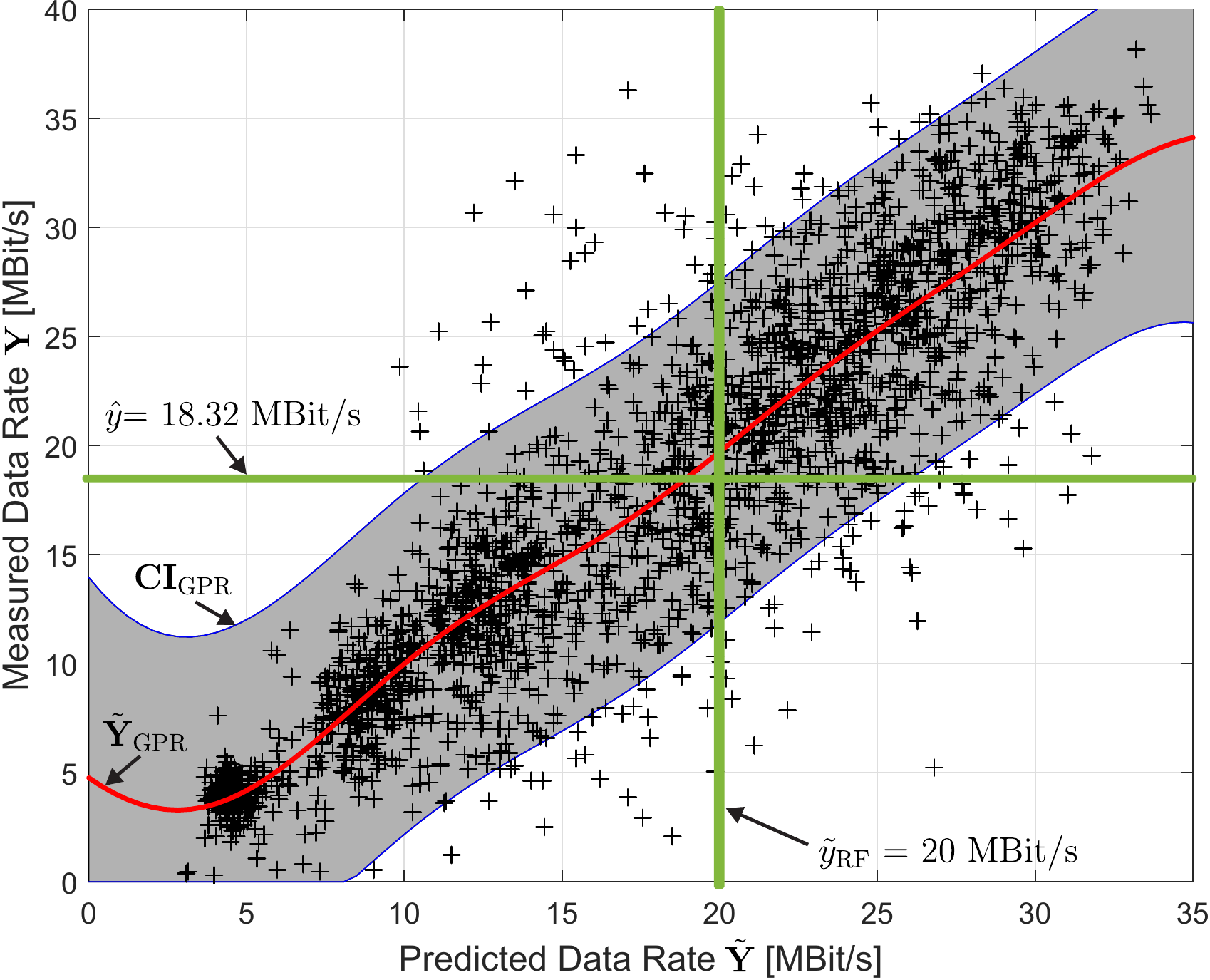}{0.48}{Usage of \ac{GPR} to derive a probabilistic description of the prediction model derivations (uplink measurements of \mnoA).}%
	\subfig{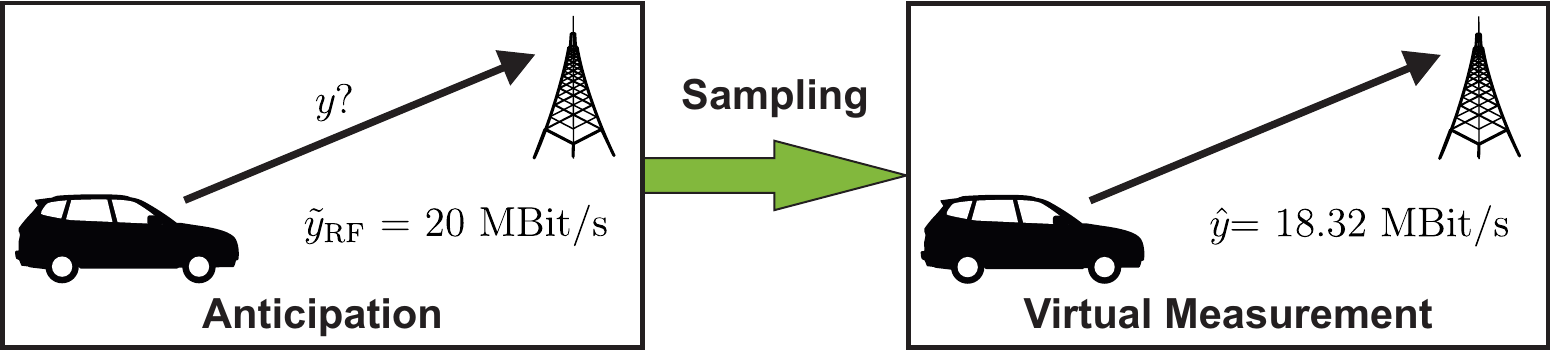}{0.48}{Application of \ac{GPR}-based derivation modeling within \ddns.}%
	
	\caption{Example behavior of the \ac{GPR} model, which is applied on the results of the regression model to transform the latter from the deterministic to the probabilistic domain. Virtual measurements are then generated by sampling from the error distribution of the predictions.}

	\label{fig:gpr_tmobile}
\end{figure}
%
%
%

%
%
\renewcommand{\sfw}{0.32}
\begin{figure*}[] 
	\centering
	
	\subfig{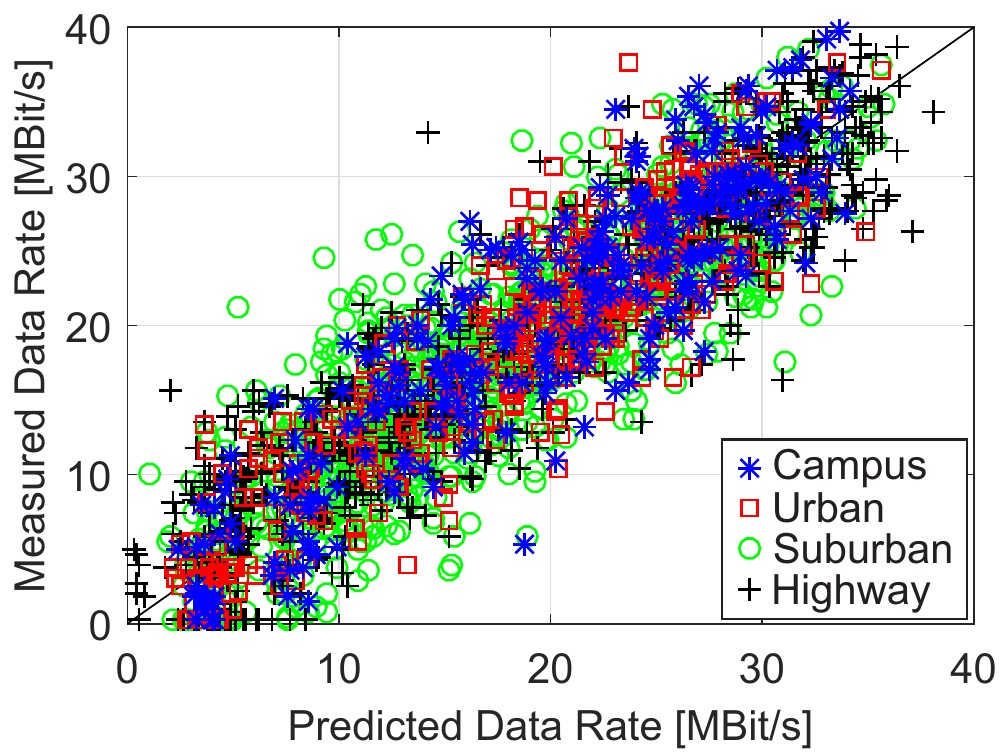}{\sfw}{MNO~A}
	\subfig{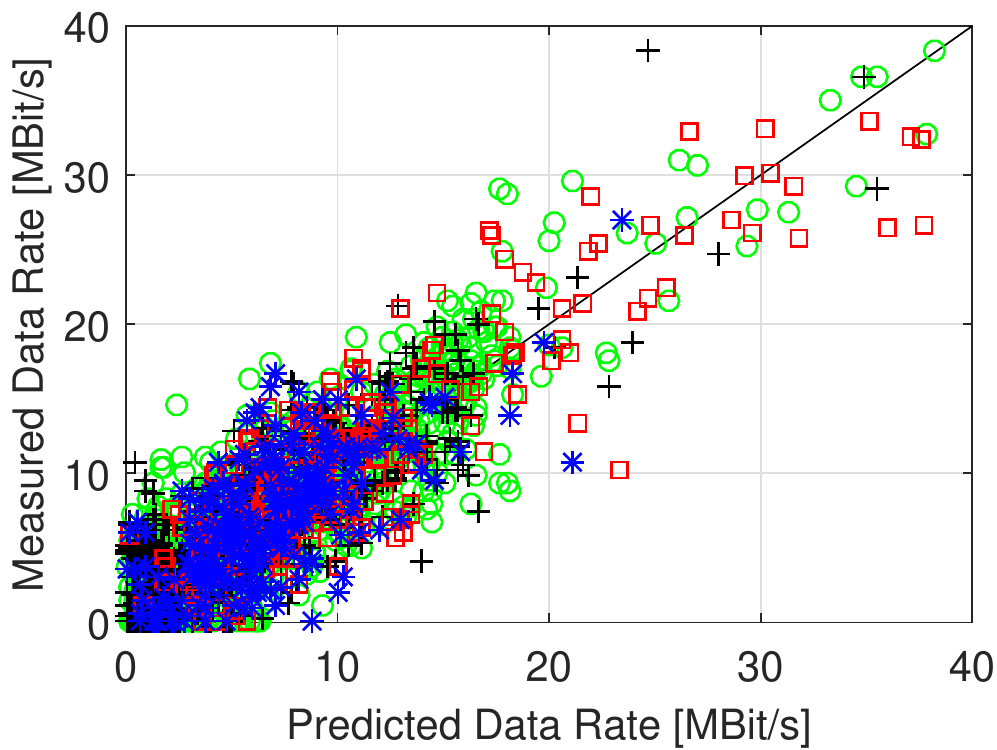}{\sfw}{MNO~B}
	\subfig{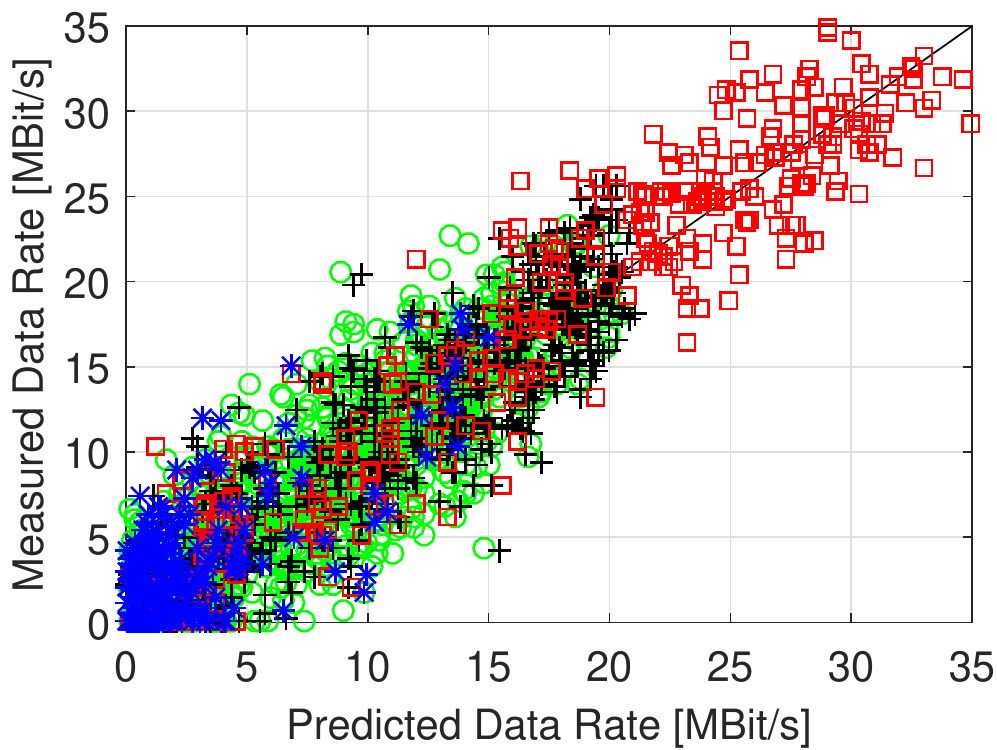}{\sfw}{MNO~C}

	\caption{Synthesized transmission profiles based on the \ac{DDNS} method by replaying the real world transmissions using \ac{RF}-based data rate prediction and \ac{GPR}-based derivation modeling (uplink transmission direction). In consideration of the real world measurements in the same scenarios (see Fig.~\ref{fig:model_performance}), it can be seen that \ac{DDNS} achieves a close to reality representation of the characteristics of all \acp{MNO}.}

	\label{fig:DDNS_distribution}
\end{figure*}
%
%
%

%
%
The deterministic data rate prediction model is now extended by a method to consider \emph{model imperfections} within the simulations in order to achieve an accurate representation of the real world behavior. Based on the analysis of the previous section, we draw the following conclusions:
%
%
\begin{itemize}
	\item In the vast majority of the evaluations, the \ac{RF} regression model achieves the most accurate prediction performance. Therefore, \ac{RF} is utilized for performing the data rate predictions within the simulation.
	\item It is more reasonable to use only few models with large data sets than a large number of highly-specified prediction models (e.g., a single model for each \ac{eNB}). Therefore, the global data sets for each \ac{MNO} and transmission direction are used as the training data for the prediction model. 
\end{itemize}
In order to derive a \emph{probabilistic} description of the derivations between ground truth and prediction model, a bayesian machine learning model is applied on the \emph{resulting transmission profile} of the prediction model. 
%
%
For this purpose, we utilize a \acf{GPR} \cite{Rasmussen/2004a} model as it inherently provides favorable statistical properties which are explained and exploited in the following paragraphs.

In the first step, the prediction results $\mYRf$ of the \ac{RF} model are used as training data for the \ac{GPR} model $f_{\text{GPR}}$ to derive a predicted data set $\mYGpr$ such that $\mYGpr = f_{\text{GPR}}(\mYRf)$.

Fig.~\ref{fig:gpr_tmobile}~(a) shows an example for the resulting behavior of the \ac{GPR} model based on the overall uplink data set of \mnoA.  
For the predicted values \yRf, the actual real world measurements are centered around \yGpr with a certain value spread. The latter describes the derivations from the real world behavior and is related to effects, which are not covered by the prediction model. However, the \emph{confidence area} of the \ac{GPR} allows to draw error-aware samples for each given value of \yRf, which follow the distribution of the real world measurements.
Assuming a gaussian distribution $\mathcal{N}$ of the prediction errors, a sample $\tilde{y}_{\text{GPR}}$ can be obtained with the standard deviation function $\boldsymbol{\sigma}_\text{GPR}$ as
%
%
\begin{equation}
	\tilde{y}_{\text{GPR}}(\tilde{y}_{\text{RF}}) 
	= \mathcal{N}\left( 
	\tilde{\mathbf{Y}}_\text{GPR}(\tilde{y}_{\text{RF}}), 
	\boldsymbol{\sigma}_\text{GPR}^2(\tilde{y}_{\text{RF}})
	\right)
\end{equation}
For the considered data set, it can be seen that that the prediction confidence is reduced for $\tilde{\mathbf{Y}}_\text{RF}<\text{3~MBit/s}$ and $\tilde{\mathbf{Y}}_\text{RF}>\text{33.5~MBit/s}$, which describes the edge regions of the training set.

%
%
Due to the probabilistic properties of the sampling process, it is possible that sample values exceed the value range of the observed measurement values or are even assigned impossible values (e.g., negative data rates). Therefore, a final filtering step is applied in order to compensate these statistical effects. The corrected sample value $\hat{y}$ is finally computed as
%
%
\begin{equation}
	\hat{y} = \begin{cases}
	\min(\mathbf{Y}_{\text{RF}}) & \tilde{y}_{\text{GPR}} < \min(\mathbf{Y}_{\text{RF}}) \\
	\max(\mathbf{Y}_{\text{RF}}) & \tilde{y}_{\text{GPR}} > \max(\mathbf{Y}_{\text{RF}}) \\
	\tilde{y}_{\text{GPR}}  & \text{else}
	\end{cases}
\end{equation}
%
%
An example application of this method within \ddns is shown in Fig.~\ref{fig:gpr_tmobile}~(b). In the \emph{anticipation} phase, the vehicle predicts the currently achievable data rate \yRf based on the passive context indicators. As a ground truth is missing in the data-driven simulation, a \emph{virtual measurement} $\hat{y}$ is derived by sampling from the confidence area of the predicted value.

%
%
For all considered \acp{MNO}, all uplink measurements were re-generated with the proposed mechanism by simulatively replaying the transmissions at their actual measurement locations under the measured network conditions. Fig.~\ref{fig:DDNS_distribution} shows the resulting distribution of \ac{DDNS}-synthesized data rate values. In comparison to the real world measurements -- see Fig.~\ref{fig:model_performance}~(a)-(c) -- it can be seen that the process is able to provide a close to reality representation of the data rate distributions, which is able to capture the \mno- as well as the scenario-specific characteristics.

	\section{Validation} \label{sec:validation}

%
%
In order to validate the proposed \ac{DDNS} method, a case study focusing on opportunistic vehicular data transfer is carried out with real world field tests serving as a ground truth. As a further reference for the performance of the proposed \ddns method, we consider classical system-level network simulation, which is based on \des. Within the simulative evaluations, both approaches replay the trajectories of the real world measurements of the highway and the suburban scenario. The ultimate goal is to \emph{mimic the real world behavior} of the analyzed anticipatory communication method within the simulation setup. The following evaluations show the results of additional validation experiments, for which the measurement data is not contained in the training sets of the machine learning methods.

%
%
It is remarked that the proposed \ddns mechanism can be exploited for catalyzing the development process of novel anticipatory networking methods by applying a method-in-the-loop approach. Within this work, the same  \texttt{C++} implementation code is used for the real world application and the \ddns variant. The only required differences are the context inputs (actual measurements in the real world, trace data in \ddns) and the data transmissions (\ac{TCP} access in the real world, machine learning-based prediction for \ddns). Achieving a similar level of \emph{code reusability} is often not possible with established network simulators, as the latter enforce the usage of simulator-specific modules and interfaces.

%
%
\renewcommand{\sfw}{0.24}
\begin{figure*}[] 
	\centering
	
	%
	%
	\subfig{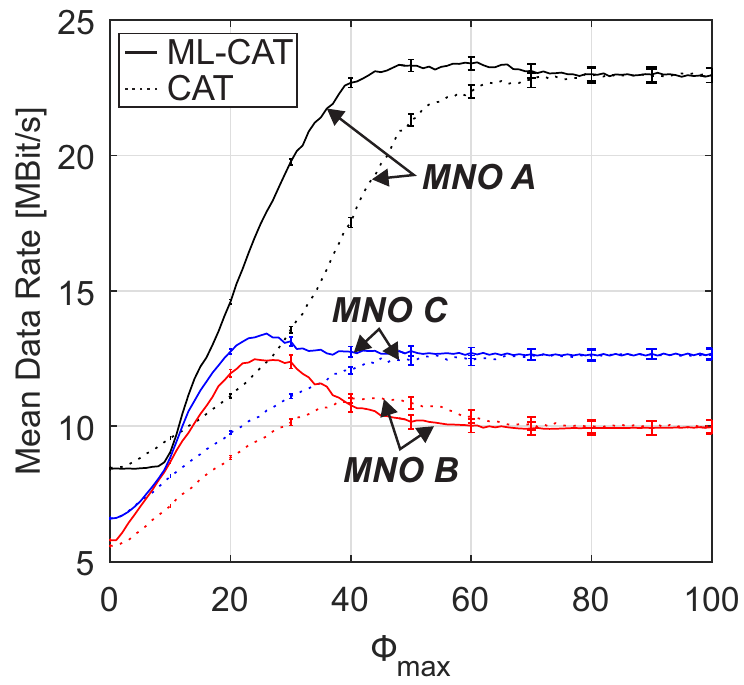}{\sfw}{\ac{CAT} and \ac{ML-CAT} (Uplink)}%
	\subfig{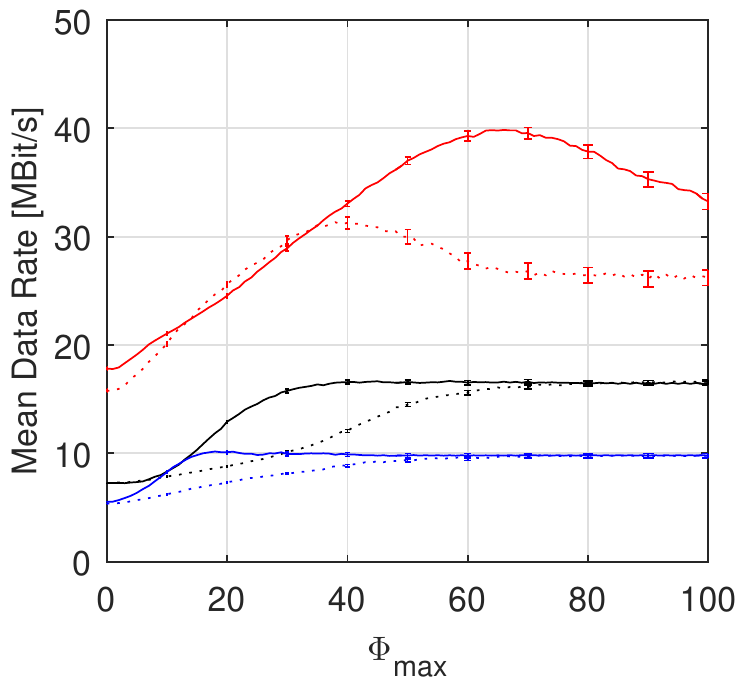}{\sfw}{\ac{CAT} and \ac{ML-CAT} (Downlink)}%
	%
	%
	\subfig{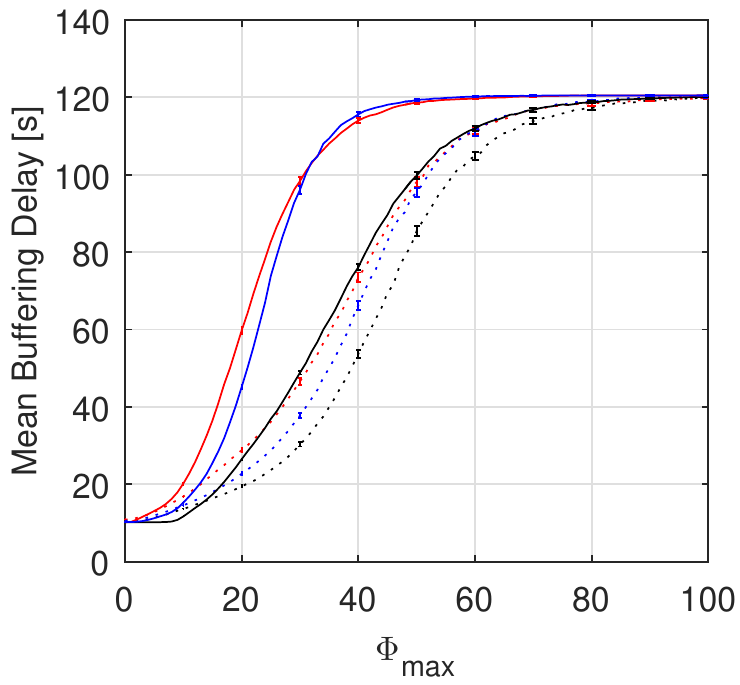}{\sfw}{\ac{CAT} and \ac{ML-CAT} (Uplink)}%
	\subfig{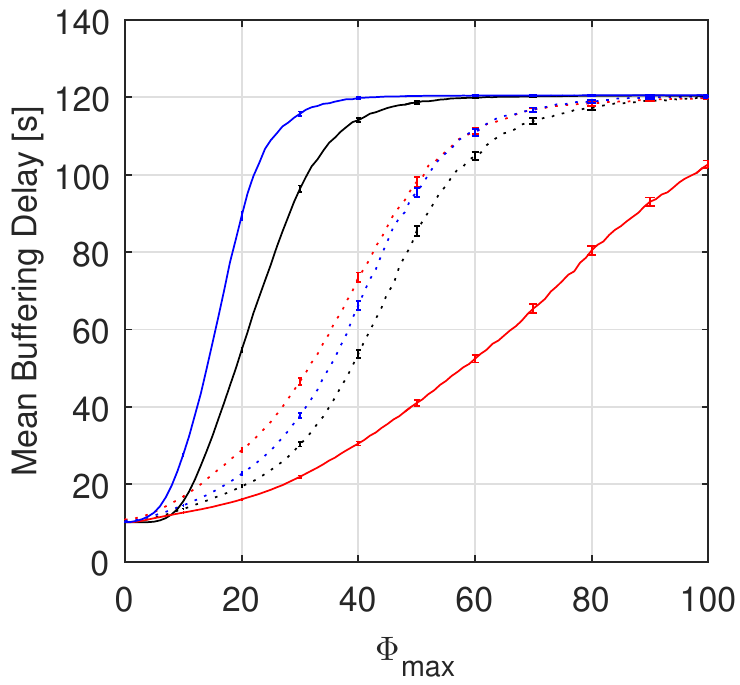}{\sfw}{\ac{CAT} and \ac{ML-CAT} (Downlink)}%
	
	\caption{\ac{DDNS}-based parameter optimization for sweet spot detection of anticipatory communication methods: Impact of the maximum metric value $\Phi_{\max}$ on data rate and buffering delay. The errorbars show the 0.95-confidence interval of the mean value. For each setup, every value of $\Phi_{\max}$ represents the aggregated performance of 500 different evaluation runs.}
	\vspace{-0.5cm}
	\label{fig:optimization}
\end{figure*}

\subsection{Anticipatory Communication Methods for Opportunistic Data Transfer} \label{sec:mus_validation}

In the following, the anticipatory communication methods, which are used as for the validation, are introduced. It is remarked that these models have been published in earlier work and are only applied here. Within this manuscript, the focus of the scientific evaluations is on the achievable accuracy of the simulation approaches and not on the performance of the transmission methods.

%
%
Within typical vehicular \ac{MTC} systems, the radio channel is accessed in a periodic way, e.g., sensor data is acquired and transmitted to a remote server with a fixed transmission interval. Since this approach does not take the current network quality into account, many transmissions are performed during low radio channel quality periods and are subject to undesired effects such as packet loss. Due to the low resulting transmission efficiency and the need for retransmissions, cell resources and energy are wasted.

%
%
In contrast to the periodic transmission approach, the considered anticipatory communication methods \textbf{\acf{CAT}} and \textbf{\acf{ML-CAT}} \cite{Sliwa/etal/2018b} access the channel in an opportunistic way based on a probabilistic process. The schemes exploit the dynamics of the network channel in the way that they delay the transmission until sufficient radio channel conditions are established. Acquired sensor data is buffered locally until a transmission decision is made for the whole buffer. Due to the introduced \emph{buffering delay}, the method is intended for delay-tolerant applications (e.g., vehicle-as-a-sensor) and does not satisfy the latency requirements of safety-critical vehicular communications.

%
%
With \textbf{\ac{pCAT}} and \textbf{\ac{ML-pCAT}} \cite{Sliwa/etal/2018a}, the general opportunistic transmission schemes are extended by a predictive component, which introduces a prediction horizon $\tau$ for forecasting the radio channel quality at the future location $\tilde{\mathbf{P}}(t+\tau)$. The latter is obtained using trajectory-aware mobility prediction and is exploited for obtaining the context data from a connectivity map.

%
%
The different \ac{CAT} variants can be configured to perform the transmission scheduling decision with respect to different metrics (e.g., \ac{SINR} and predicted data rate). In the first step, the measured metric value $\Phi(t)$ is transformed to a normed metric value $\Theta(t)$ with
%
%
\begin{equation}
	\Theta(t) =  \frac{\Phi(t)-\Phi_{\min}}{\Phi_{\max}-\Phi_{\min}} 
\end{equation}
in order to allow the application of the basic \ac{CAT} principles with metrics that have different value ranges $[\Phi_{\min}, \Phi_{\max}]$. The transmission probability $p_{\text{TX}}(t)$ is then computed as
%
%
\begin{equation}
	p_{\text{TX}}(t) = \begin{cases}
	0 & \Delta t < t_{\min} \\
	1 & \Delta t > t_{\max} \\
	\Theta(t)^{\alpha\cdot z}  & \text{else} \\
	\end{cases}
\end{equation}
with $\alpha$ being an exponent, which describes how much the scheme should prefer high metric values and $ \Delta t$ being the passed time since the last transmission has been performed. $t_{\min}$ is used to guarantee a minimum payload size and $t_{\max}$ defines an upper bound for the buffering delay.
$z$ is a \ac{pCAT}-exclusive factor, which is responsible for taking the trade-off between the current measurement $\Phi(t)$ and the anticipated future network quality $\tilde{\Phi}(t+\tau)$ into account and is computed as 
%
%
\begin{equation}
	z = \begin{cases}
	\max(|\Delta \Phi(t) \cdot (1-\Theta(t)) \cdot \gamma)|, 1) & \Delta \Phi(t) > 0 \\
	(\max(|\Delta \Phi(t) \cdot \Theta(t) \cdot \gamma)|, 1))^{-1} & \Delta \Phi(t) \leq 0 
	\end{cases}
\end{equation}
with $ \Delta \Phi(t)  = \tilde{\Phi}(t+\tau) - \Phi(t)$ and a prediction weighting factor $\gamma$. The probabilistic transmission decision process itself is triggered periodically (1~Hz in the following evaluations).

\subsection{Reference Setup for System-level Network Simulation}

%
%
As a reference for the methodological evaluation, a classical system-level network simulation approach based on \des is applied with \ac{OMNeT++}~5.0 \cite{Varga/Hornig/2008a}, INET~3.4 and SimuLTE~v0.9.1 \cite{Virdis/etal/2015a}. The provided example scenario \texttt{test\_handover} is taken as a starting point for own extensions.
As pointed out in Sec.~\ref{sec:introduction}, multiple simplifications are required for transforming the real world scenario into a system-level simulation setup:
%
%
\begin{itemize}
	\item \textbf{Code extension}: SimuLTE uses a single carrier frequency definition for all \acp{eNB} within a scenario. Therefore, the simulator implementation was extended to support individual carrier frequencies for each \ac{eNB} according to their corresponding real world values.
	\item \textbf{Unknown \ac{MNO} configuration}: Within the real world, the resource scheduling mechanisms are \mno-specific and unknown for the client devices. SimuLTE implements proportional fair scheduling which might differ from the mechanisms used by the considered \acp{MNO}.
	\item \textbf{Simplified prediction model}: As the simulator only models a fraction of the features of the prediction model -- which is used by the metrics of  ML-CAT and ML-pCAT -- a reduced version of the latter needs to be applied within the simulative evaluation. For each \mno, a machine learning-based prediction model is trained using the payload size, \ac{SINR} and frequency features for uplink and downlink direction. Considering the feature importance analysis in Sec.~\ref{sec:featureImportance}, it can be concluded that multiple important impact factors are omitted with these simplifications.
	\item \textbf{Missing features}: Since the applied transmission power of the \enb is unknown, the SimuLTE default value is applied for all base stations. In addition, there are no implementations for \ac{CA} and for the \ac{TPC} mechanism of the \ac{UE}.
\end{itemize}
%
%
%
%
%
\newcommand{\entry}[2]{&#1 & #2 \\}
\newcommand{\head}[2]{& \toprule \entry{\textbf{#1}}{\textbf{#2}}}

\renewcommand{\baselinestretch}{1}

\begin{table}[ht]
	\small
	\centering
	\caption{General parameters of for the validation.}
	\begin{tabular}{p{0.2cm}p{3.5cm}p{3.8cm}}
		\toprule
		\sideHeader{11}{0.5cm}{General} 
		\entry{\textbf{Parameter}}{\textbf{Value}}
		\midrule
		\entry{Data source}{50~kByte/s}
		\entry{Evaluation interval}{1~Hz}
		\entry{$t_{\min}$}{10~s}
		\entry{$t_{\max}$}{120~s}
		\entry{$\alpha$}{6}
		\entry{$\gamma$ (pCAT)}{2}
		\entry{$\gamma$ (ML-pCAT)}{0.5}

		\midrule
		
		\sideHeader{9.5}{0.3cm}{SimuLTE} 
		\entry{Carrier frequency}{$\lbrace900,1800,2100\rbrace$~MHz}
		\entry{Bandwidth}{20~Mhz}
		\entry{\ac{UE} transmission power}{23~dBm}
		\entry{\acs{eNB} transmission power}{43~dBm}
		\entry{Channel model}{WINNER II Urban Macro}
		\entry{Other parameters}{\texttt{test\_handover} defaults}

		\bottomrule
		
	\end{tabular}
	\label{tab:parameters}
\end{table}

\ifdefined\singleColumn
	\renewcommand{\baselinestretch}{1.5}
\else
\fi

In the following result analysis, the SimuLTE evaluations will be referred to as \emph{\des}. Tab.~\ref{tab:parameters} summarizes the overall parameterization of the transmission schemes and the \des configurations.
%
%
\renewcommand{\entry}[4]{#1 & #2 & #3 & #4 \\}
\renewcommand{\head}[4]{\toprule \entry{\textbf{#1}}{\textbf{#2}}{\textbf{#3}}{\textbf{#4}}\midrule}
\renewcommand{\cW}{0.4cm}

\aboverulesep = 1pt
\belowrulesep = 1pt

\renewcommand{\baselinestretch}{1.0}

\begin{table}[ht]
	\small
	\centering
	\caption{Parameterization of the opportunistic transmission schemes.}
	\begin{tabular}{p{2.8cm}|p{\cW}p{\cW}|p{\cW}p{\cW}|p{\cW}p{\cW}}
		\midrule
		 & \multicolumn{2}{c|}{\textbf{\mno A}} & \multicolumn{2}{c|}{\textbf{\mno B}} & \multicolumn{2}{c}{\textbf{\mno C}} \\
		\textbf{$\Phi_{\max}$} & \textbf{UL} & \textbf{DL} & \textbf{UL} & \textbf{DL} & \textbf{UL} & \textbf{DL} \\

		\midrule
		
		CAT [dB] & 30 & 30 & 30 & 30 & 30 & 30 \\
		pCAT [dB] & 30 & 30 & 30 & 30 & 30 & 30 \\
		ML-CAT [MBit/s] & 30 & 30 & 20 & 50 & 20 & 15 \\
		ML-pCAT [MBit/s] & 30 & 30 & 20 & 50 & 20 & 15 \\
		
		\bottomrule

	\end{tabular}

	\label{tab:metrics}
	\vspace{0.1cm}
	\emph{UL}: Uplink, \emph{DL}: Downlink
\end{table}

\ifdefined\singleColumn
	\renewcommand{\baselinestretch}{1.5}
\else
\fi

\subsection{\ac{DDNS}-based Parameter Optimization}

%
%
\renewcommand{\sfw}{0.49}
\begin{figure*}[] 
	\centering
	
	\subfig{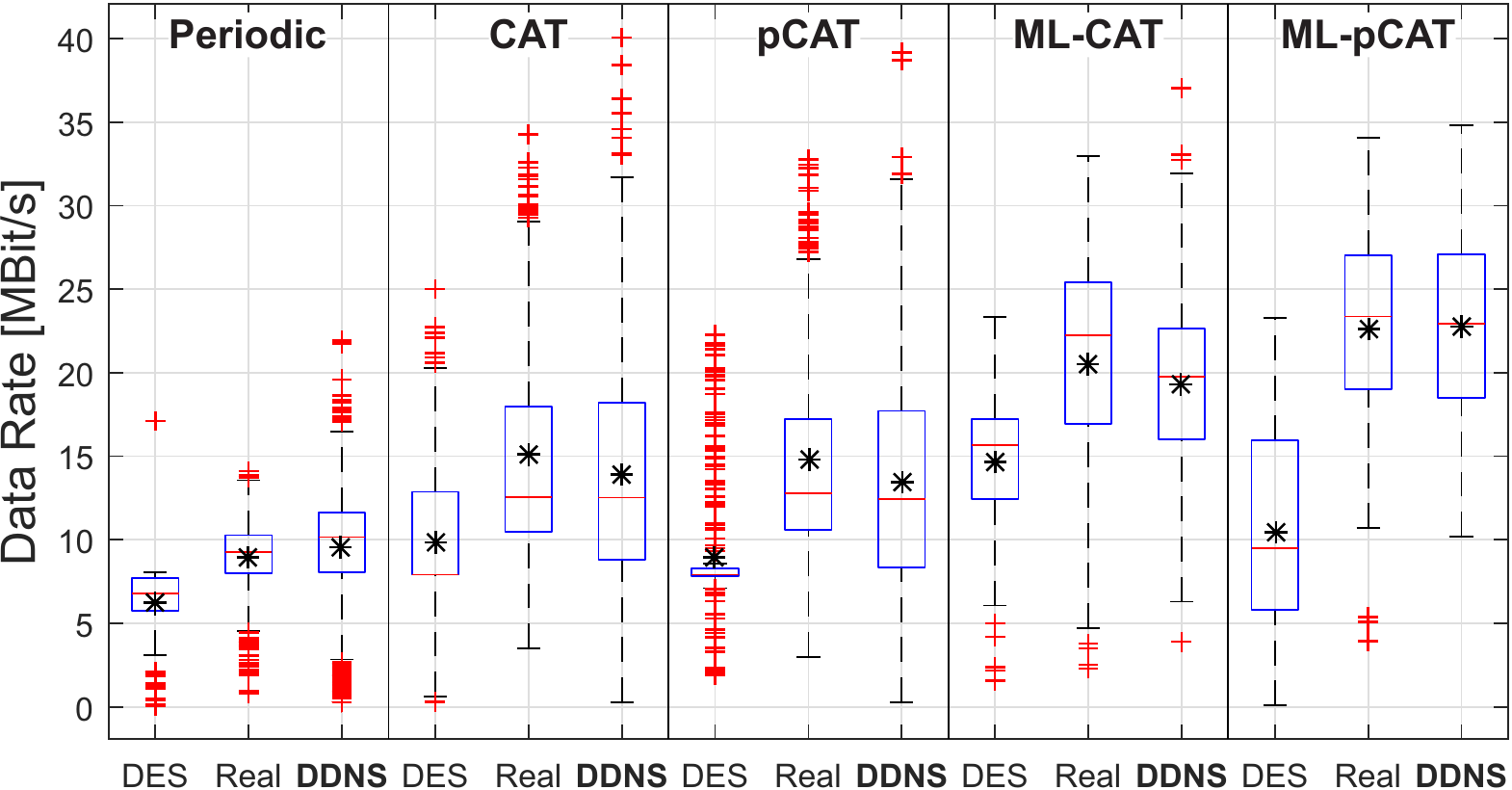}{\sfw}{MNO~A (Uplink)}
	\subfig{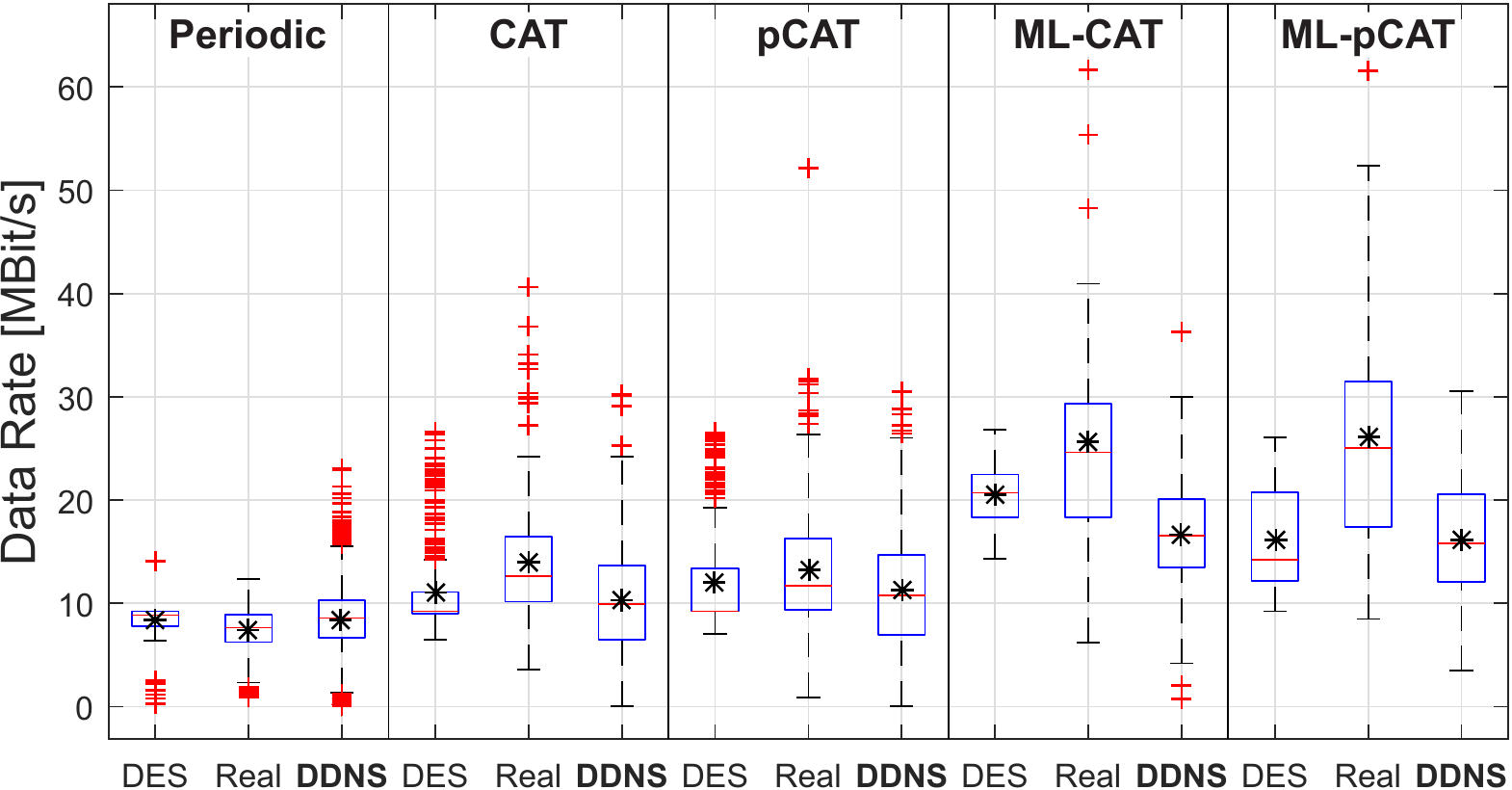}{\sfw}{MNO~A (Downlink)}
	
	\vspace{-0.2cm}

	\subfig{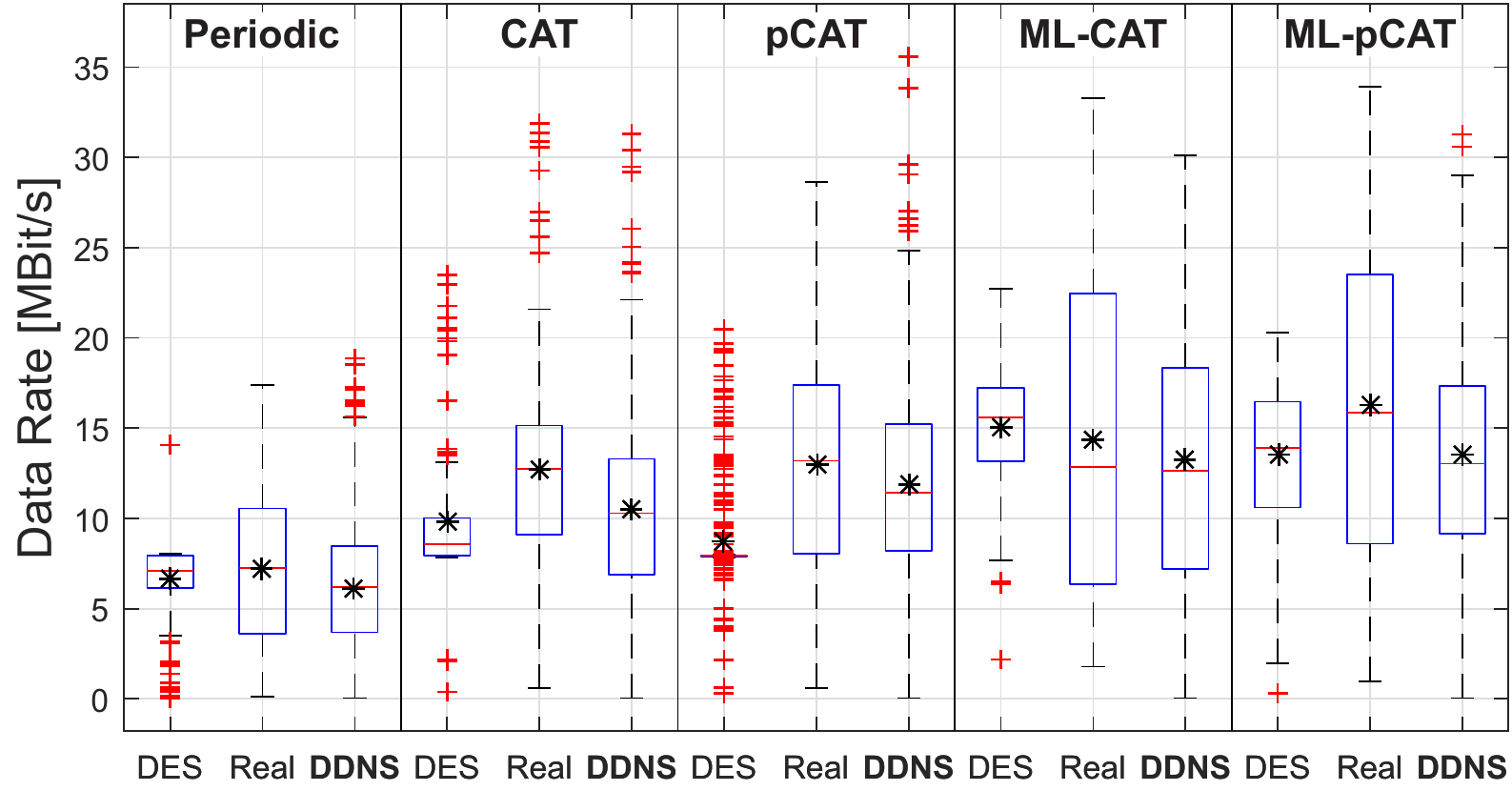}{\sfw}{MNO~B (Uplink)}
	\subfig{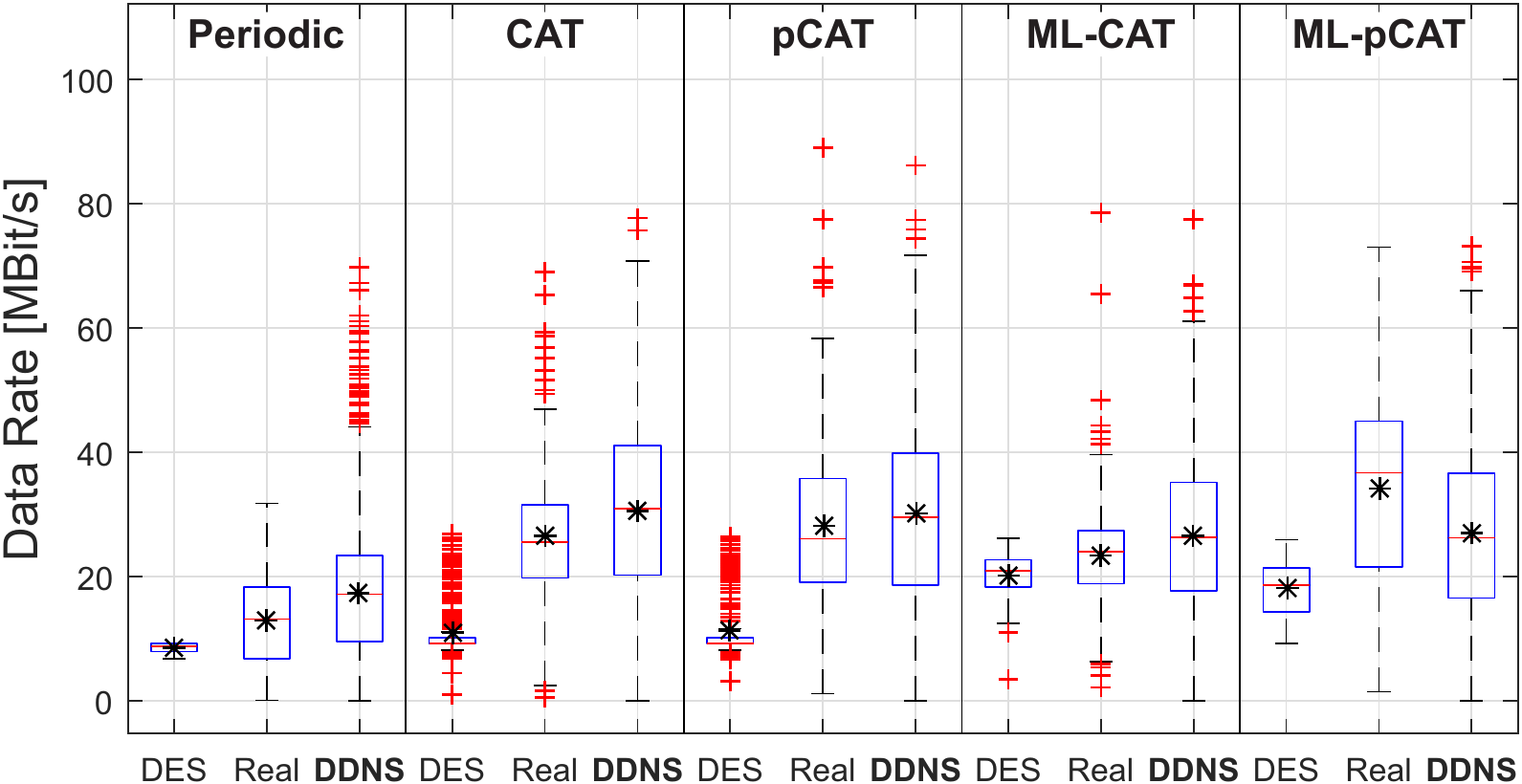}{\sfw}{MNO~B (Downlink)}
	
	\vspace{-0.2cm}

	\subfig{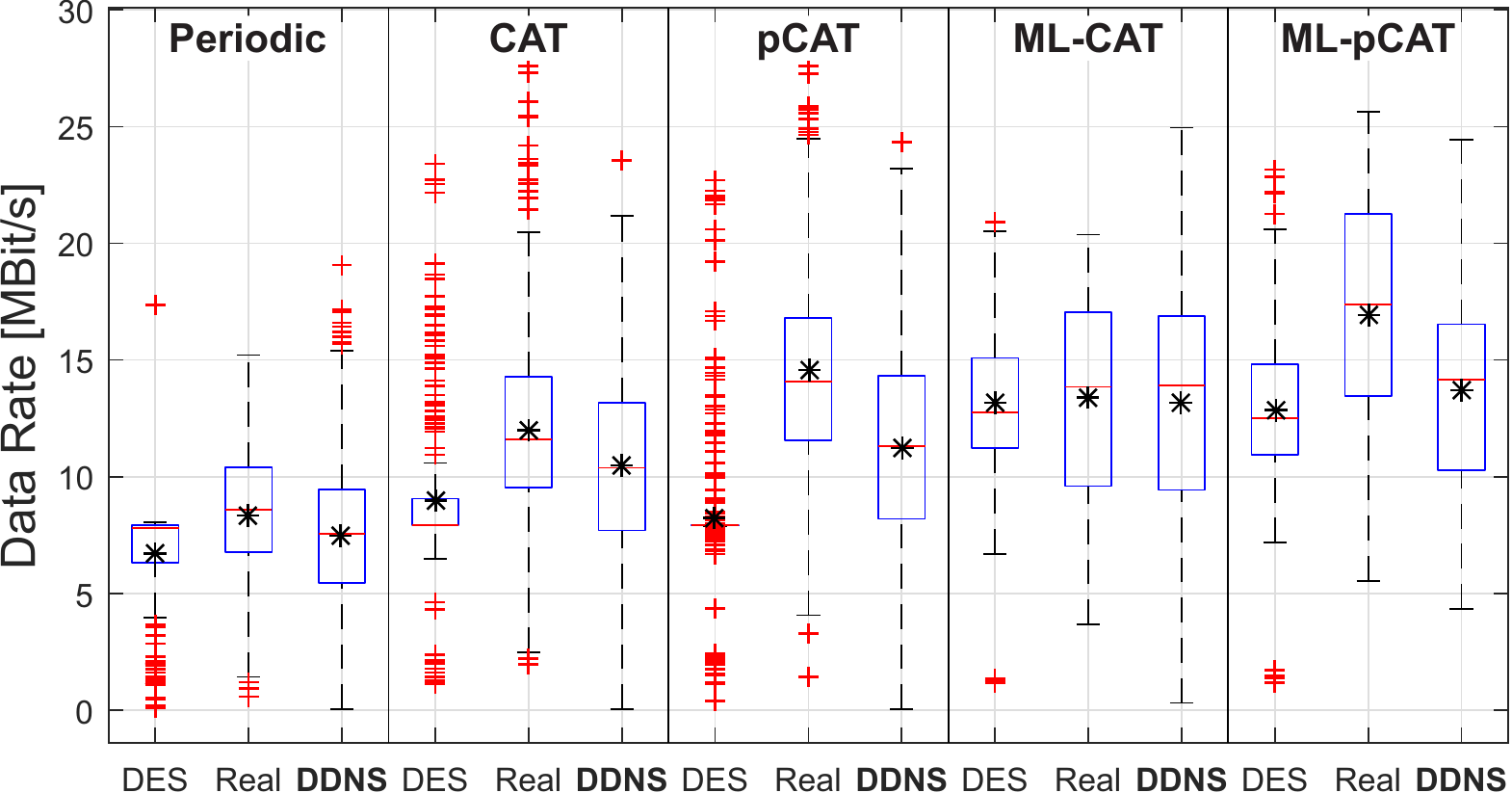}{\sfw}{MNO~C (Uplink)}
	\subfig{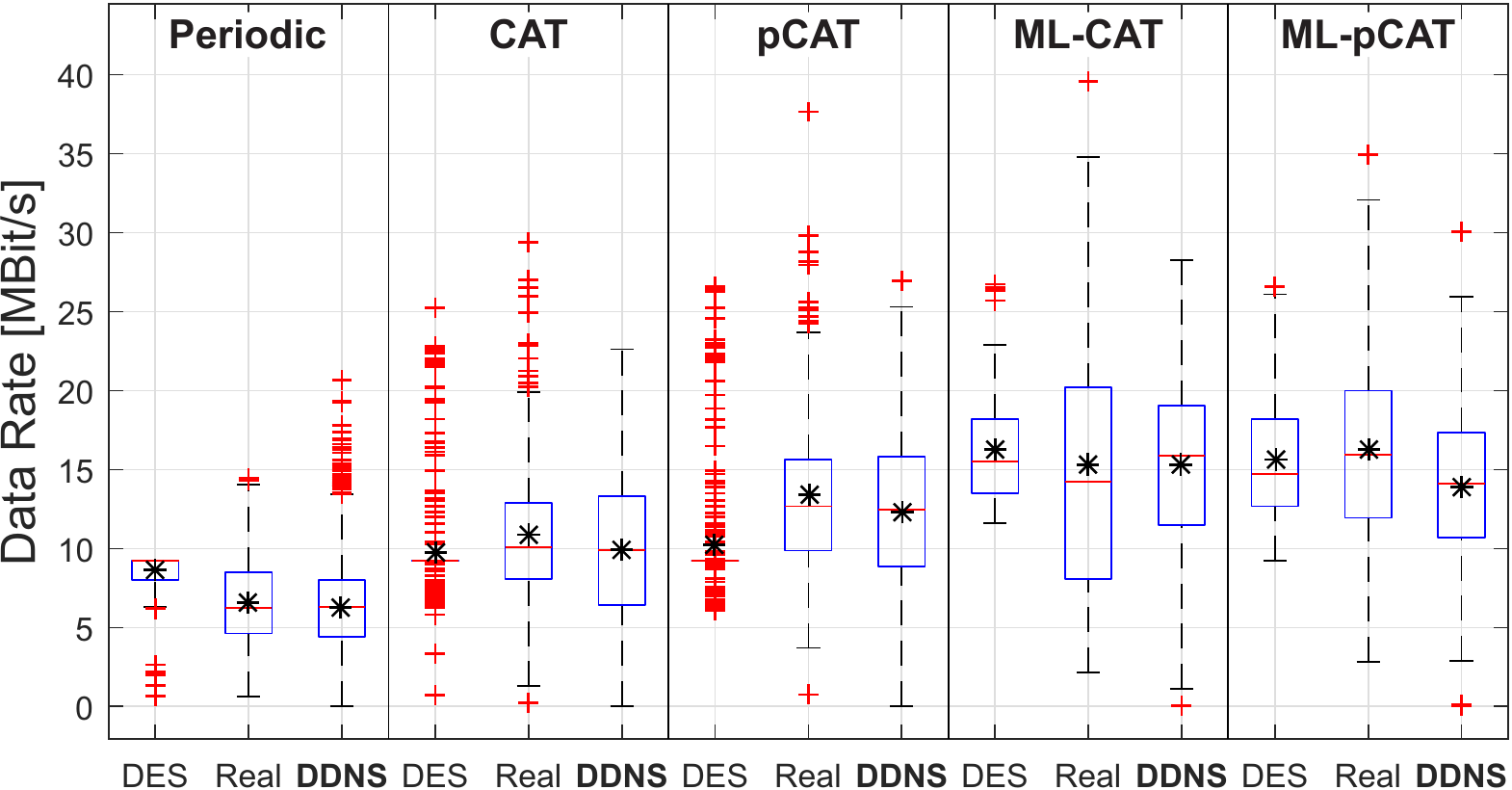}{\sfw}{MNO~C (Downlink)}
	
	\caption{Comparison of the resulting end-to-end behavior of the different transmission schemes for the considered evaluation methods in uplink and downlink direction. The goal of the \des and \ddns methods is to mimic the real world behavior of the different data transfer methods. The real world results consist of additional data, which was obtained exclusively for the validation and is not contained in the training sets of the machine learning models. For a summary of the key findings, see Fig.~\ref{fig:ecdf_coeff}.}
	\label{fig:cat_validation}
\end{figure*}

%
%
Since \ac{DDNS} evaluations can be performed in a highly resource-efficient way (see Sec.~\ref{sec:scalability}), even large-scale parameter studies that employ \emph{brute-force} analysis over the whole parameter space can be executed.
%
%
In order to find the best parameterizations for the considered anticipatory communication methods for each \mno and transmission direction, the impact of $\Phi_{\max}$ on the average resulting data rate and buffering delay is analyzed in Fig.~\ref{fig:optimization}. 
As extremely high metric values (e.g., $\text{SINR}>50~\text{dB}$) do not occur in the real world data set, the transmission schemes converge as they are determined by the maximum buffering delay $t_{\max}$ which enforces the transmissions after exceeding the timeout.

%
%
Every parameter configuration is evaluated based on 20 mobility traces and each evaluation is repeated with 25 different random seeds. In total, for each \mno, every transmission scheme is analyzed in 50000 different evaluation runs. It is obvious, that performing the same amount of evaluations in the real world is practically impossible as it would imply to analyze the data transfer schemes on a total driven distance of more than 1.15 million~km during more than 950 whole days. Classical system-level network simulation would take more than 2600~days with four computation cores (estimated based on the findings in Sec.~\ref{sec:scalability}). However, the \ac{DDNS} approach requires less than three hours to finish on the considered evaluation system.

Tab.~\ref{tab:metrics} shows the resulting \mno-specific parameterization of the different transmission schemes, which is based on a trade-off between data rate and buffering delay. Note that the units for $\Phi_{\max}$ differ between the transmission schemes, as \ac{CAT} and \ac{pCAT} perform their decisions with respect to the measured \ac{SINR}, while \ac{ML-CAT} and \ac{ML-pCAT} consider the predicted data rate of the \ac{RF} model. For all schemes, $\Phi_{\min}$ is configured as the zero value of the corresponding unit. As a reference, periodic data transfer with a fixed interval of 10~s is considered.

\subsection{Resulting Modeling Accuracy} \label{sec:modeling_accuracy}

Finally, the resulting modeling accuracy is investigated for system-level network simulation and the proposed \ac{DDNS}.
Fig.~\ref{fig:cat_validation} shows the resulting end-to-end data rate values for the different transmission schemes and \acp{MNO} in uplink and downlink direction.
%
%
Within the real world evaluation, several characteristics by applying the different \ac{CAT} variants can be observed:
\begin{itemize}
	\item The periodic transmission scheme provides the lower baseline for the achievable data rate as the transmissions are performed unaware of the network channel conditions.
	\item The \ac{SINR}-based \ac{CAT} variants are able to increase the resulting data rate significantly.
	\item With the introduction of machine learning-based channel quality assessment (ML-CAT) the average data rate is massively increased.
	\item By using context-prediction (\ac{pCAT} and \ac{ML-pCAT}), an additional slight improvement is achieved.
\end{itemize}
%
%
For \ddns, the achievable modeling accuracy is directly related to the prediction accuracy of the applied regression models (see. Fig.~\ref{fig:model_performance}). Therefore, \mnoA achieves a significantly more realistic representation of the real world behavior for the uplink than for the downlink.
%
%
As \ac{ML-CAT} utilizes data rate prediction within the transmission scheme itself, it is subject to the accumulated error of the \ddns mechanism and the prediction error of the $\Phi_{RF}$ metric within the \ac{CAT} mechanism itself. \ac{ML-pCAT} is furthermore impacted by the prediction error for the anticipated data rate at the future location $\tilde{\mathbf{P}}(t+\tau)$.
However, in the vast majority of all evaluations, the impact of the aggregated prediction errors has a lower impact on the results than the parameter uncertainties of the \des.

%
%
A general observation for the \des results is that the different \mnos behave very similar. As the \mno-specific configurations are unknown, the \mnos only differ with respect to the \enb position and the applied carrier frequencies. However, in the real world and in the \ddns evaluations, different behavioral characteristics for the \mnos can be observed. Although \mnoA achieves the highest uplink data rates for all transmission schemes in the real world, it has the lowest throughput in the event-based simulation. Due to the high prediction accuracy in the uplink for \mnoA, \ac{ML-pCAT} is able to unleash its full potential in the real world and in the \ddns evaluation, where it achieves an average data rate gain of $\sim14$~MBit/s. However, this effect is not captured by the \des due to the applied simplified regression model and the missing \ac{TPC}. Contrastingly, it shows a similar behavior as for the other \mnos.
%
%
\mnoB achieves the highest mean data rate in the downlink by applying \ac{CA} within some of the cells. As this feature is not explicitly modeled within the SimuLTE framework, the observed behavior differs significantly from the real world. In contrast to that, the proposed \ac{DDNS} approach is able to implicitly learn the \emph{impacts of \ac{CA}} on the considered \ac{KPI} directly from the measurement data.

%
%
It can be seen that the \des fails to mirror the real world behavior of the \ac{pCAT} transmission scheme. Due to the context prediction step, \ac{pCAT} is highly sensible to the \ac{SINR} dynamics. In the \des, the network dynamics differ from the real world due to the fixed \enb transmission power of 43~dBm. In the real world, \enb position and transmission power optimization are the results of a complex \emph{network planning} phase, which is performed with respect to the radio environment. In contrast to that, the proposed \ddns does not require definitions or value assumptions for the \enb parameters, it simply learns the implications of the hidden variable on the considered end-to-end \ac{KPI}.

%
%
\basicFig{}{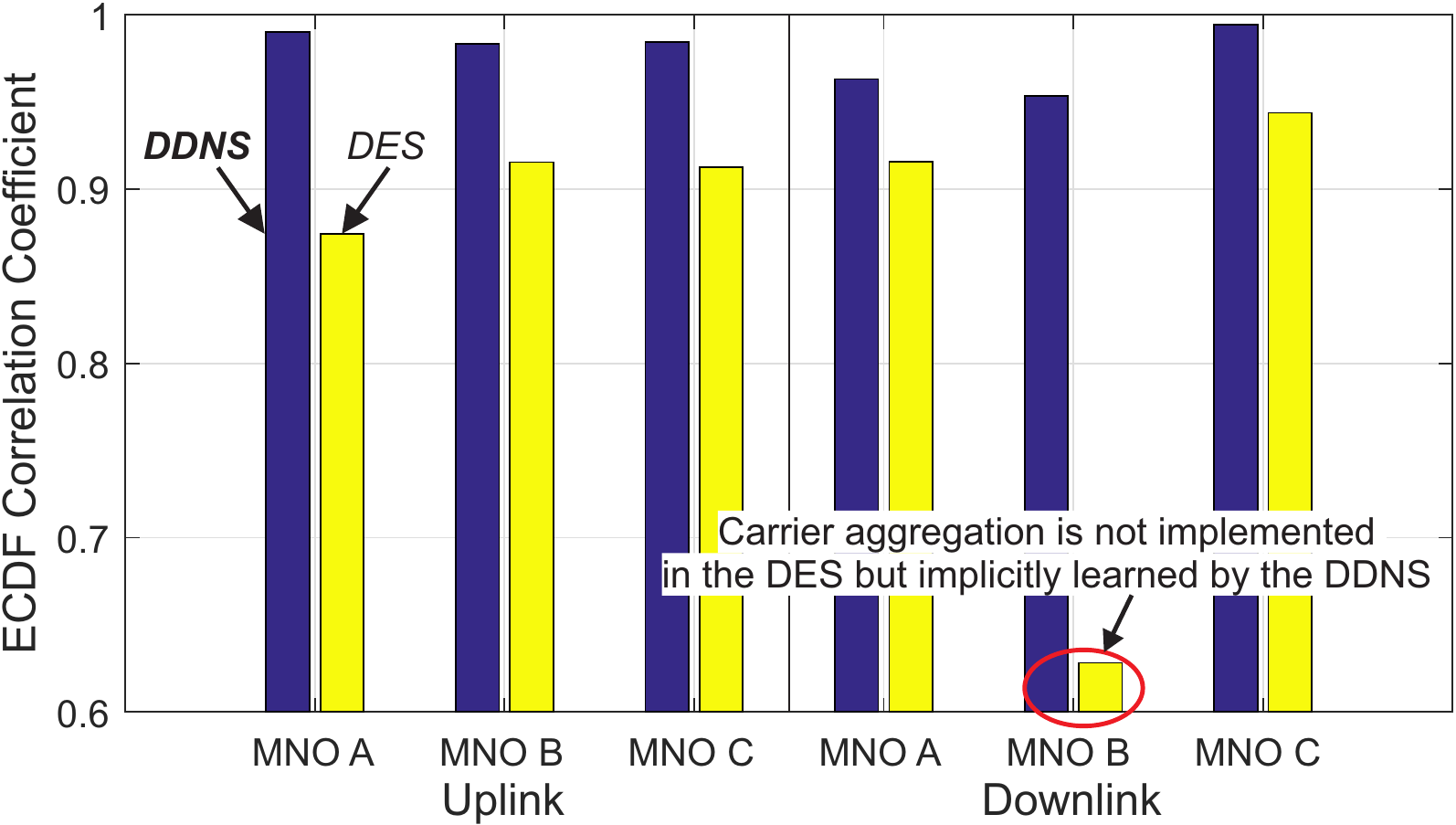}{Overall similarity between real world behavior and simulation models. The bars shows the correlation coefficient of the \acf{ECDF} of \des and the proposed \ddns method with the empirical measurements.}{fig:ecdf_coeff}{-0.5cm}{0cm}{1}
In order to assess the overall similarity between real world and simulation, for each transmission scheme, a similarity measurement is computed as the correlation coefficient of the \acp{ECDF} of the real world measurement results and the corresponding simulation results. Fig.~\ref{fig:ecdf_coeff} summarizes the average behavior of \ddns and \des. It can be seen that the proposed \ddns achieves a significantly higher modeling accuracy than the \des method in all considered cases.

\subsection{Computational Efficiency} \label{sec:scalability}

In additional to the achievable modeling accuracy of the obtained results, the computational efficiency of the simulation setup itself is of great importance for the system optimization phase.
%
%
\basicFig{}{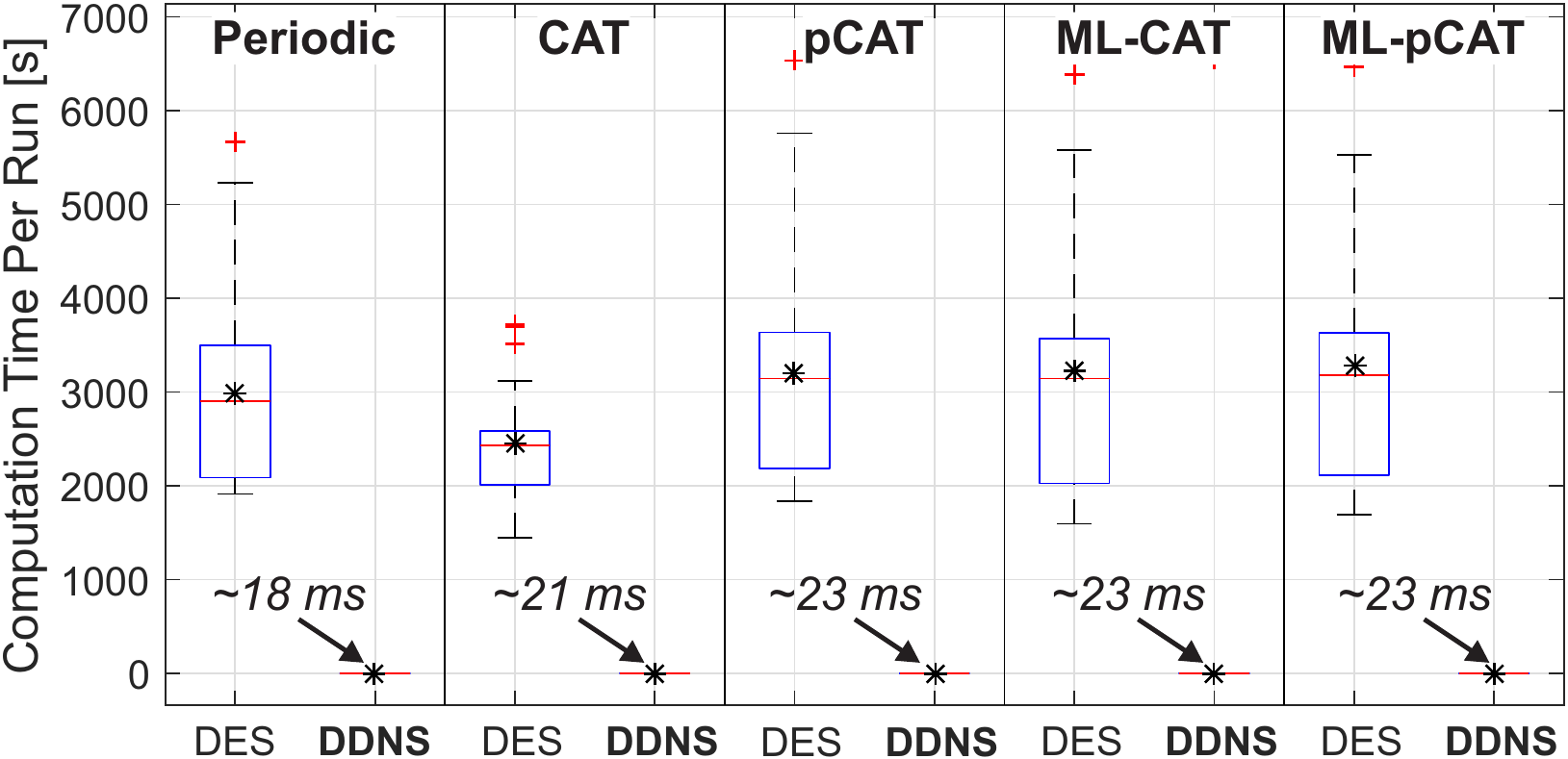}{Comparison of the average computation times per scenario evaluation for the proposed \ac{DDNS} method and an established system-level network simulation setup.}{fig:computation_times}{-0.5cm}{0cm}{1}
Fig.~\ref{fig:computation_times} shows the aggregated resulting computation time per run for the different evaluations methods for the considered transmission schemes. It can be seen that the proposed \ddns is multiple orders of magnitude faster than the \des approach. Although the application-level end-to-end behavior of the data transfer method is investigated, the \des spends most of its computation resources on simulating processes that are only indirectly related to the considered \ac{KPI}. 
%
%
As an example, within the SimuLTE setup, neighboring \acp{eNB} are interconnected based on \emph{X2} interfaces in order to coordinate the cellular handover mechanisms which is completely simulated during the evaluations. In consequence, the event-based network simulation does not scale well when the number of \acp{eNB} is increased.
Contrastingly, the proposed \ddns allows to derive results with a very high computational efficiency as the machine learning-based modeling focuses on the end-to-end behavior itself and treats the intermediate modules as a black box. 

	\section{Limitations of Data-driven Network Simulation} \label{sec:limitations}

%
%
Although the previous evaluations have pointed out numerous advantages of using the \ddns method for analyzing end-to-end network performance indicators, it needs to be remarked that the proposed method has a defined application range with specific limitations.

\begin{itemize}
	\item \textbf{Dependency to the prediction model:}
	%
	Since the acquired real world data provides the foundation for the evaluation scenario and the prediction models, the significance of the \ddns results is severely depending on the quality and the amount of the data (see Sec.~\ref{sec:rf_behavior}). 
	%
	%
	Due to the focus on analyzing end-to-end indicators in a data-driven way, the considered features need to be carefully chosen in the data acquisition phase. In contrast to system-level network simulation, it is mostly not possible to alter the analyzed \ac{KPI} without performing additional measurements and model trainings.

	\item \textbf{Scenario-oriented analysis}:
	Replaying real world context traces allows to analyze the performance of new data transfer methods under close to reality network conditions. However, the results are only significant for the considered evaluation scenarios and  the existing configurations of the network infrastructure. Although this limits the generalizability of the achieved results, it needs to be remarked that system-level network simulators are confronted with the same issues. In addition, the latter are further impacted by simulator-specific feature derivations (e.g., models implemented in \emph{\ac{DES}~A} might be missing in \emph{\ac{DES}~B}) which limit the significance of cross-simulator performance comparisons \cite{Cavalcanti/etal/2018a}.
	%
	%
	Open data sets serving as \emph{reference scenarios} could make a significant contribution to improving the generalizability of the \ac{DDNS} approach. This way, a novel method could be evaluated using a wide range of different \ac{MNO}- and scenario- specific impact factors. 	

	\item \textbf{Black box approach:}
	%
	Although the applied black box approach enables very fast result generation, the implied encapsulation does not allow to inspect the behavior of the intermediate layers. Therefore, \ddns is mainly intended to be used as a powerful method for the system optimization phase, when the most important features and indicators have already been explored.
	However, for analyzing the behavior of the lower layer protocols, existing end-to-end models for these layers (eg., \cite{Ye/etal/2018b, Doerner/etal/2018a, Aoudia/Hoydis/2018a}) can be applied in a similar way. A possible future extension might be a hierarchical \ddns setup, where the prediction models of the upper layers leverage the results of the lower layer prediction models as additional features.
\end{itemize}

\section{Conclusion} \label{sec:conclusion}

%
%
In this paper, we presented \acf{DDNS} as a novel methodological approach for analyzing anticipatory vehicular communication systems. The proposed method exploits machine learning-based prediction models and crowdsensing-enabled data acquisition for achieving close to reality modeling of end-to-end network performance indicators.

%
%
While classic \des-based system-level network simulation suffers from a high scenario generation complexity due to a large number of parameters uncertainties, \ddns is able to learn their hidden interdependencies implicitly solely from real world measurement data.
%
%
The statistics of the derivations between prediction model and real world behavior can be learned by a dedicated machine learning model in order to consider their implications as gaussian noise within the simulative evaluation phase.

%
%
Applying \ddns to model the behavior of cellular communication systems requires to train individual models for each \mno. Although machine learning-based data rate prediction is able to consider the effects of cross-layer dependencies, the resulting end-to-end behavior is significantly depending on unknown \mno-specific configurations (e.g., the resource scheduling mechanisms).

%
%
As it was shown in the proof-of-concept validation focusing on anticipatory vehicular data transmission, the proposed \ddns method is able to achieve more realistic end-to-end results with a significantly higher computational efficiency than the reference system-level network simulation setup.  

%
%
In future work, we want to further exploit crowdsensing-based data maintenance for keeping the simulation data consistent with the real world. By introducing online learning capabilities in the regression phase, an up-to-date digital twin of the real world network could be achieved, which would be able to autonomously learn and consider new technological developments (similar to \ac{CA} as discussed in Sec.~\ref{sec:modeling_accuracy}).

	\renewcommand{\baselinestretch}{1}
	\section*{Acknowledgment}

\footnotesize
Part of the work on this paper has been supported by Deutsche Forschungsgemeinschaft (DFG) within the Collaborative Research Center SFB 876 ``Providing Information by Resource-Constrained Analysis'', project B4.
		
	\bibliographystyle{IEEEtran}
	\bibliography{Bibliography}
	
	\normalsize
\begin{IEEEbiography}[{\includegraphics[width=1in,height=1.25in,clip,keepaspectratio]{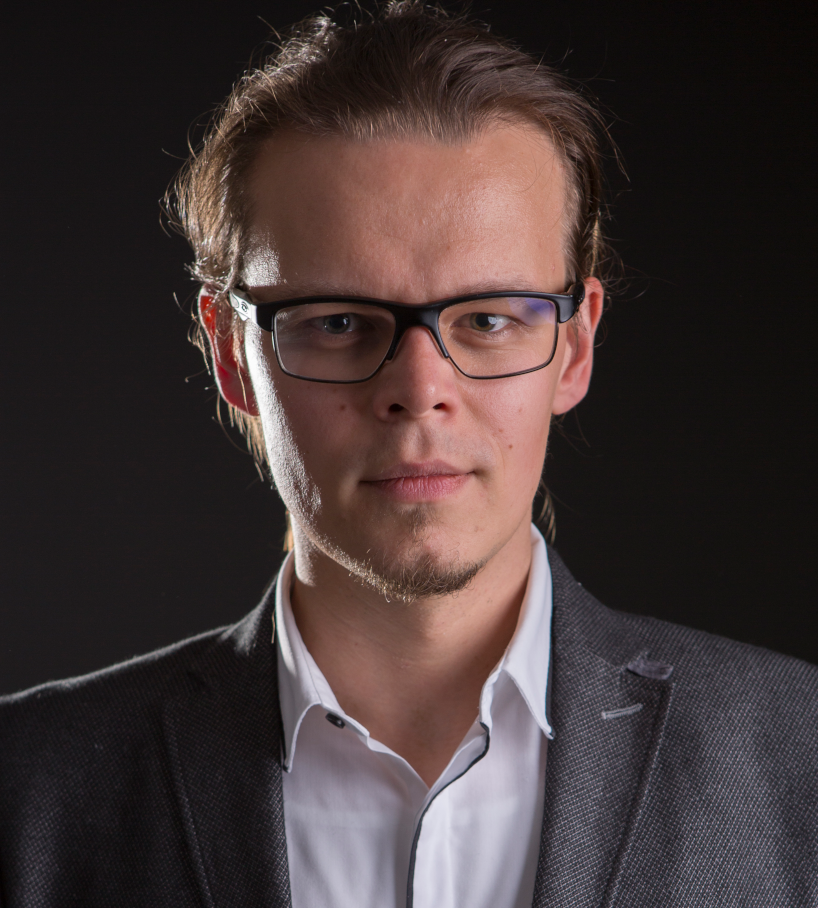}}]
{Benjamin Sliwa}
(S'16) received the M.Sc. degree from TU Dortmund University, Dortmund, Germany, in 2016. He is currently a Research Assistant with the Communication Networks Institute, Faculty of Electrical Engineering and Information Technology, TU Dortmund University. He is working on the Project "Analysis and Communication for Dynamic Traffic Prognosis" of the Collaborative Research Center SFB 876. His research interests include predictive and context-aware optimizations for decision processes in vehicular communication systems. Benjamin Sliwa has been recognized with a Best Student Paper Award at IEEE VTC-Spring 2018 and the 2018 IEEE Transportation Electronics Student Fellowship "For Outstanding Student Research Contributions to Machine Learning in Vehicular Communications and Intelligent Transportation Systems".
\end{IEEEbiography}

\begin{IEEEbiography}
[{\includegraphics[width=1in,height=1.25in,clip,keepaspectratio]{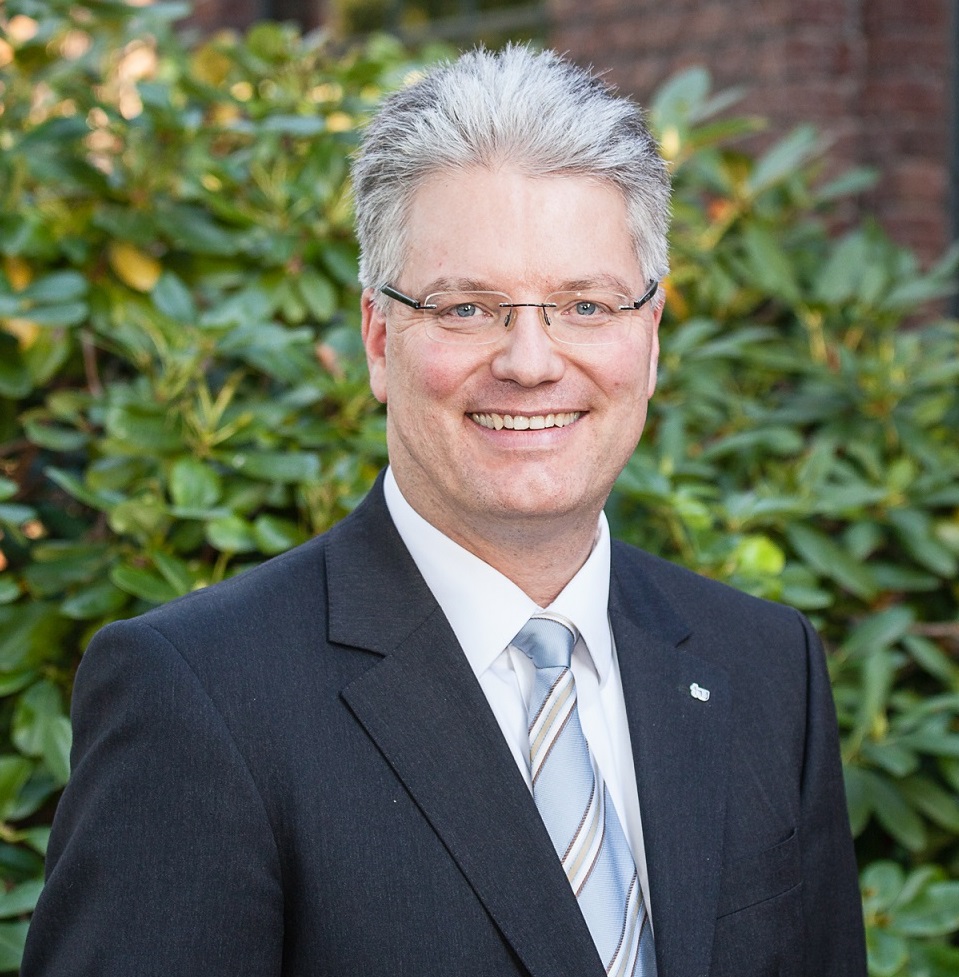}}]
{Christian Wietfeld}
(M'05-SM'12) received the Dipl.-Ing. and Dr.-Ing. degrees from RWTH Aachen University, Aachen, Germany.  He is currently a Full Professor of communication networks and the Head of the Communication Networks Institute, TU Dortmund University, Dortmund, Germany. For more than 20 years, he has been a coordinator of and a contributor to large-scale research projects on Internet-based mobile communication systems in academia (RWTH Aachen '92-'97, TU Dortmund since '05) and industry (Siemens AG '97-'05). His current research interests include the design and performance evaluation of communication networks for cyber-physical systems in energy, transport, robotics, and emergency response.  He is the author of over 200 peer-reviewed papers and holds several patents. Dr. Wietfeld is a Co-Founder of the IEEE Global Communications Conference Workshop on Wireless Networking for Unmanned Autonomous Vehicles and member of the Technical Editor Board of the IEEE Wireless Communication Magazine. In addition to several best paper awards, he received an Outstanding Contribution award of ITU-T for his work on the standardization of next-generation mobile network architectures.

\end{IEEEbiography}

\end{document}